\documentclass[10pt,preprint]{aastex}

\newcommand{\e}{\epsilon}

\newcommand{\gp}{\gamma^\prime}
\newcommand{\Dp}{\Delta^\prime}
\newcommand{\el}{\ell_{\rm S}}
\newcommand{\pp}{p^\prime}
\newcommand{\Vp}{V^\prime}
\newcommand{\Np}{N^\prime}

\newcommand{\tp}{t^\prime}

\newcommand{\psim}{\lower.5ex\hbox{$\; \buildrel \propto \over\sim \;$}}

\shorttitle{Particle Acceleration in Gamma Ray Bursts}
\shortauthors{Dermer and Humi}

\begin{document}
\title{Adiabatic Losses and Stochastic Particle Acceleration \\
    in Gamma-Ray Burst Blast Waves}

\author{Charles D.  Dermer\altaffilmark{1} and Mayer Humi\altaffilmark{2,1}}

\altaffiltext{1}{E.  O.  Hulburt Center for Space Research, Code 7653,
Naval Research Laboratory, Washington, DC 20375-5352}
\altaffiltext{2}{Mathematical Sciences Department, Worcester Polytechnic Institute, 100 Institute Road, Worcester, MA 01609-2280}

\begin{abstract}

We treat the problem of adiabatic losses and stochastic particle acceleration in gamma-ray burst (GRB) blast waves that decelerate by sweeping up matter from an external medium. The shocked fluid is assumed to be represented by a homogeneous expanding shell. The energy lost by nonthermal particles through adiabatic expansion is converted to the bulk kinetic energy of the outflow, permitting the evolution of the bulk Lorentz factor $\Gamma$ of the blast wave to be self-consistently calculated.  The behavior of the system is shown to reproduce the hydrodynamic self-similar solutions in the relativistic and nonrelativistic limits, and the formalism is applicable to scenarios that are intermediate between the adiabatic and fully radiative regimes. 

Nonthermal particle energization through stochastic gyroresonant acceleration with magnetic turbulence in the blast wave is treated by employing energy-gain rates and diffusive escape timescales based upon expressions derived in the quasilinear regime. If the magnetic field in the shocked fluid approaches its equipartition value, this process can accelerate escaping particles to $\gtrsim 10^{20}$ eV energies, consistent with the hypothesis that ultra-high energy cosmic rays (UHECRs) are accelerated by GRB blast waves. Due to particle trapping by the magnetic turbulence, only the highest energy particles can escape during the prompt and afterglow phases of a GRB for acceleration by a Kolmogorov spectrum of MHD turbulence. Lower energy particles begin to escape as the blast wave becomes nonrelativistic and shock Fermi acceleration becomes more important. 

\end{abstract}


\section{Introduction}

The origin of cosmic rays (CRs) with energies above the knee of the CR spectrum at $\approx 3\times 10^{15}$ eV is not established. Nor is the origin of ultra-high ($\gtrsim 10^{19}$ eV) energy cosmic rays (UHECRs) known. Because of their large Larmor radii, UHECRs probably originate from extragalactic sources. Theoretical difficulties \citep{lc83} to accelerate cosmic rays to energies above the knee energy via shock Fermi acceleration in supernova remnants suggests that a new class of sources with adequate power and acceleration efficiency is required. The stellar progenitors of gamma-ray bursts (GRBs) provide a plausible solution \citep{mu95,mu96, wax95,vie95}. One motivation for this proposal is the coincidence \citep{wax95,vie95} between the power supplied by GRB sources within the Greisen-Zatsepin-Kuzmin photopion energy-loss radius and the power needed to produce the observed energy density of UHECRs. This coincidence has recently been verified \citep{der00} for the external shock model of GRBs \citep{bd00}, under the assumption that the rate density of GRBs follows the star formation history of the universe. A second motivation is that relativistic shock waves provide a site to accelerate particles to much higher energies than is possible in nonrelativistic shock waves. For example, reflection from a relativistic shock with Lorentz factor $\Gamma$ will increase a particle's energy by a factor $\sim \Gamma^2$. GRB blast waves with $100\lesssim \Gamma \lesssim 1000$ can therefore accelerate particles to energies near the knee of the cosmic ray spectrum in a single cycle, with subsequent acceleration producing cosmic rays above the knee energy \citep{vie95,mu96}. 

GRB emission originates either from internal shocks in a relativistic wind \citep{rm94} or from an external shock \citep{rm92,mr93} that forms when a relativistic blast wave decelerates and is energized by sweeping up matter from the surrounding medium. Acceleration to energies above the knee of the CR spectrum depends on the specific acceleration mechanism. \citet{gal99} show that an external shock model to accelerate UHECRs starting from nonrelativistic particles is not viable. In this paper, we instead consider a scenario for UHECR acceleration based upon stochastic particle acceleration in GRB blast waves that slow from relativistic to nonrelativistic speeds \citep{wax95,rm98,sd00}. Although we focus on gyroresonant acceleration in relativistic shocks formed by the sources of GRBs, this approach is also applicable to stochastic acceleration in systems with mildly relativistic and nonrelativistic shocks. For example, Type Ib/c SNe produce shocks with speeds that often exceed $0.1c$ \citep{wei00}. Mildly relativistic shocks with Lorentz factor $\Gamma \sim 1.6$-2 were produced by the Type Ic SN 1998bw \citep{kul98}. Irrespective of whether GRB 980425 is associated with SN 1998bw \citep{pia99}, the existence of SN 1998bw demonstrates that unusual SNe such as SN 1998bw eject relativistic flows that will contribute to CR particle acceleration. The adiabatic and stochastic processes considered here will also play a role in particle acceleration in SNRs, although the first-order systematic energy-gain rate dominates that of stochastic gyroresonant acceleration in nonrelativistic shocks.

In Section 2, we treat adiabatic losses in GRB blast waves, and provide an equation for blast-wave evolution that accounts for these losses in a manner that self-consistently rechannels the energy lost from particle expansion into the directed kinetic energy of the expanding blast wave. An accurate treatment of adiabatic losses is required to estimate maximum particle energies through stochastic acceleration. We assume throughout that the blast wave is uniform and can be described by a thin shell of outflowing matter, whereas in reality the flow is subject to instabilities \citep{kjr92} and nonlinear feedback from the accelerated particle distribution \citep{be87,bar99} that can smooth the transition at the shock front. Acoustic instabilities can also cause a turbulent fragmentation of the shocked fluid \citep{rv87}. Although the turbulence so generated can be effective in generating a magnetic field near equipartition with the downstream flow, it also invalidates the simple assumption of a uniform shocked shell, thus limiting the applicability of the method. Future work must consider how instabilities and nonlinear effects of the shock structure will modify the results presented here.

Stochastic particle acceleration through gyroresonant acceleration is considered in Section 3 by generalizing quasilinear expressions to a fully turbulent regime with relativistic Alfv\'en speeds, as probably applies to GRB blast waves. We show that UHECR acceleration is possible in relativistic blast waves by considering various limits to particle acceleration, including a comparison of the particle gyroradius with the blast wave width \citep{hil84}, and competition with adiabatic losses, synchrotron losses, and diffusive escape. Our work improves upon previous treatments that have made similar comparisons \citep{vie95,vie98a,rm98}, though here we restrict our considerations to stochastic particle acceleration in GRB blast waves. Particle energy evolution and escape from the blast wave is examined through the use of a particle continuity equation. 

Simulations of the spectra of particles accelerated in GRB blast waves are presented in Section 4. Particles accelerated through stochastic processes in relativistic blast waves will also participate in shock Fermi processes as the blast wave decelerates to nonrelativistic speeds, so that a full calculation of the spectrum of the accelerated particles must consider both first- and second-order Fermi acceleration. GRB sources may thus be the sources of UHECRs and CRs above the  knee of the CR spectrum, as well as contributing some portion of lower energy CR hadrons. Summary and conclusions are given in Section 5.
 
\section{Evolution of Blast Wave Lorentz Factor}

Both the particles that are captured from an external medium by a GRB blast wave and the ejected thermal plasma particles that initially carry most of the explosion energy will be subject to adiabatic losses due to volume expansion of the blast wave shell. This energy is reconverted into the directed kinetic energy of the outflow. Here we present a treatment of expansion losses in a blast wave that contains particles with arbitrary energies in the comoving frame. The treatment is valid for a blast wave with general values of the bulk speed, and is therefore applicable to both decelerating relativistic GRB blast waves and to nonrelativistic SNR shock waves. The treatment applies to general deceleration regimes from the fully adiabatic to the fully radiative limit.

\subsection{Blast-Wave Equation of Motion}

Let $\pp = \sqrt{\gamma^{\prime2}-1}$ represent the dimensionless momentum of a particle with Lorentz factor $\gp$, where primes refer to quantities in the comoving frame. The particle distribution function $\Np (\pp;x)$, integrated over the volume of the blast wave shell, is defined so that $\Np (\pp;x)d\pp = [d\Np (\pp;x)/d\pp]d\pp$ represents the differential number of particles with momenta between $\pp$ and $\pp + d\pp$ in a blast wave at radius $x$. In writing this function, we assume that the particles are isotropically distributed in the comoving frame. Global conservation of energy implies that 
\begin{equation}
d(\Gamma U_{tot})= dm + \Gamma dU_{rad}\; .
\label{GUtot}
\end{equation}
All masses are in energy units. Here $\Gamma$ is the Lorentz factor of the blast wave, $dm$ is the differential change in the swept-up rest-mass energy, $U_{tot}$ is the total internal energy in the comoving frame, and $dU_{rad}$ is the differential change in the comoving internal energy that is radiated from the blast wave. The radiated energy is assumed to be isotropically emitted in the comoving frame. We do not take into account energy gains when the system scatters or captures external radiation or magnetic-field energy. Thus equation (\ref{GUtot}) can apply to internal emission processes such as synchrotron, synchrotron self-Compton, bremsstrahlung, photomeson and secondary nuclear production processes, but not to Compton scattering processes involving external photon fields. 

The quantity
\begin{equation}
U_{\rm tot} = \mu m_p\int_0^\infty dp^\prime \gp \Np (\pp;x) \equiv M + U  =  M_0 + m(x) + U\end{equation}
represents the total internal energy, including rest-mass energy, whereas 
\begin{equation}
U = \mu m_p\int_0^\infty dp^\prime (\gp-1) \Np (\pp;x) \label{Utot}
\label{U}
\end{equation}
represents the internal energy with rest-mass energy excluded. The total rest mass $M$ consists of the sum of the mass $M_0 = E_0/\Gamma_0$ ejected by the explosive event with energy $E_0$ and initial Lorentz factor $\Gamma_0$, in addition to the swept-up mass $m(x) = \int_0^x d\tilde x \; [dm(\tilde x)/d\tilde x] = 4\pi \mu m_p \int_0^x d\tilde x \tilde x^2 n_{ext}(\tilde x)$. Here $n_{ext}(x)$ is the density of the external medium, and is assumed for simplicity to have radial symmetry. The quantity $\mu$ is the ratio of the mass of swept-up particles to the mass of swept-up protons, assuming that the different ionic species have the same comoving distribution function. In most case, the protons make the dominant contribution to the swept-up nonthermal particle energy; thus $\mu \cong 1$.  Generalization to multiple species of swept-up particles, including leptons, ions and charged dust \citep{sd00} is straightforward but is not treated here. The explosion is assumed to be uncollimated, though it is also straightforward to generalize to a system with beamed energy releases and to external media with arbitrary density distributions. 

The internal energy $U$ in the blast wave increases by the addition of the kinetic energy of the swept-up matter, and decreases due to particle energy losses through adiabatic expansion and radiation.  Thus $dU = dU_m + dU_{adi} + dU_{rad}$, where $dU_m =(\Gamma - 1)dm$ and $dU_{adi}$ represent the changes in internal energy due to swept-up particle kinetic energy and adiabatic losses, respectively.  Expanding equation (\ref{GUtot}) gives $d[\Gamma(U+M)] = \Gamma dU + \Gamma dm +(U+M)d\Gamma$. From this follows the equation of blast wave evolution
\begin{equation}
-\;{d\Gamma\over \Gamma^2 - 1} = {dm + ({\Gamma\over P^2}) dU_{adi}\over M+U}\;,
 \label{dGdx}
\end{equation}
where the blast wave momentum $P = \sqrt{\Gamma^2 -1}$. Equation (\ref{dGdx}) can equivalently be written as
\begin{equation}
-\;{dP\over dx} = {P\Gamma(dm/dx) + ({\Gamma^2\over P}) (dU_{adi}/dx)\over M_0+m(x)+U}\;.
\label{dPdx}
\end{equation}
The second term in the numerator on the right-hand side of either equations (\ref{dGdx}) or (\ref{dPdx}) represents the impulse to the blast wave resulting from adiabatic energy losses of the particles in the blast wave.  Equation (\ref{dGdx}) generalizes the equation derived by \citet{pmr98} for an adiabatic blast wave by the inclusion of radiative losses in the calculation of $U$ and $dU_{adi}/dx$.

For a strongly radiative blast wave, the adiabatic loss term is small and the radiative equation of blast-wave evolution derived by \citet{bm76} is recovered by setting $U = dU_{adi}/dx = 0$ and $dm = dM$, giving
\begin{equation}
-\;{d\Gamma \over \Gamma^2 -1} = {dm \over M_0 +m(x)}\;
\label{Grad}
\end{equation}
This is easily solved to obtain
\begin{equation}
\Gamma(x) = {[1+m(x)/M_0]^2(\Gamma_0+1) + \Gamma_0 - 1 \over  [1+m(x)/M_0]^2(\Gamma_0+1) - \Gamma_0 +1} \; .
\label{Grad_sol}
\end{equation}

When radiation losses are included but adiabatic losses are neglected, equation (\ref{dGdx}) reduces to the form 
\begin{equation}
-\;{d\Gamma\over \Gamma^2 - 1} = {dm \over M_0+m(x)+U}\;
\label{dGdxmom}
\end{equation}
that is derived from a momentum- and energy-conservation analysis \citep{dc98,cd99,pir99}. This equation can be solved \citep{cd99} by noting that $dU = (\Gamma - 1)dm$. Furthermore, if a fraction $\epsilon$ of the swept-up kinetic energy is instantaneously radiated from the blast wave, then $dU = (1-\e)(\Gamma - 1) dm$. Defining ${\cal M} = M_0 + m(x) + U$ implies that $d{\cal M} = [\e +\Gamma(1-\e )]dm$, one obtains 
\begin{equation}
-\;{d\Gamma \over \Gamma^2 -1} = {d{\cal M} \over {\cal M}[\e +\Gamma(1-\e )]  }\; .
\label{Gmom}
\end{equation}

Equation (\ref{Gmom}) has been used to treat partially radiative blast waves \citep{pir99,bd00,msb00}.  \citet{hdl99,hdl00} suggest using the relation $d{\cal M} = dm[2\Gamma(1-\epsilon)+\epsilon]$ so that the blast-wave evolutionary behavior also follows the Sedov behavior in the adiabatic nonrelativistic regime. Both approaches are adequate to examine the approximate behavior of relativistic blast waves in intermediate radiative regimes, and the treatment of Huang and colleagues furthermore approximates the correct dependence of  nonrelativistic adiabatic blast waves. But these approaches do not properly treat adiabatic losses of particles within the blast wave. Moreover, it is necessary to make the assumption that $\epsilon \approx$ const throughout the regime of blast-wave evolution under consideration. When questions of particle acceleration are treated, it is essential to have a correct treatment of adiabatic losses, which equation (\ref{dPdx}) provides. 

\subsection{Adiabatic Particle Losses}

If the internal energy $U$ in the blast wave is dominated by the kinetic energy of a thermal particle distribution, then $U$ changes due to volume expansion according to the relation
$U^{-1}(dU_{adi}/dx) = - (\hat \gamma - 1)(d\ln \Vp / dx)$, where $\hat \gamma$ is the ratio of specific heats, and $\hat \gamma = 4/3$ and $5/3$ for relativistic and nonrelativistic monatomic gases, respectively, and $\Vp$ is the comoving fluid volume. The general case involving a mixed fluid that includes both thermal and nonthermal particles can be derived by noting that $\pp$ changes through expansion losses according to the expression
\begin{equation}
-\;({d\pp\over dx})_{adi} =  {\pp\over 3} {d\ln\Vp\over dx}\;.
\label{dpdx}
\end{equation}
Equation (\ref{dpdx}) reproduces the limiting adiabatic loss forms exhibited by monatomic thermal gases, noting that in the relativistic limit $\pp \gg 1$,  $d\gp \rightarrow -(\gp/3)d\ln\Vp$, and in the nonrelativistic limit $\pp \ll 1$, $d(p^{\prime 2}/2)\rightarrow -(2/3)(p^{\prime 2}/2)d\ln\Vp$. 

The comoving volume of the fluid shell is approximated by the expression
\begin{equation}
\Vp = 4 \pi x^2 \Delta^\prime = 4\pi x^2 ({f_\Delta x\over \Gamma})\;,
\label{Vp}
\end{equation}
where $\Delta^\prime$ is the comoving width of the blast-wave shell. As noted in the Introduction, this is an extreme simplification due to instabilities and turbulence that will modify the shell structure \citep{rv87,mrp94}. When $f_\Delta = 1/12$, this relation is consistent with the approximation $\Vp \cong m(x)/(n^\prime m_p)$ \citep{pmr98} in the limits $\Gamma \gg 1$ and $\Gamma -1 \ll 1$, where $n^\prime (x) = (\hat \gamma \Gamma + 1) n_{ext}/(\hat\gamma -1)$ is the downstream comoving density in terms of the external medium density $n_{ext}$. 
The differential adiabatic expansion energy-loss rate for particles in a GRB blast wave is therefore given by
\begin{equation}
({d\pp\over dx})_{adi} = -\; \pp ({1\over x} - {1\over 3}{d\ln \Gamma\over dx})\;,
\label{dpdxadi}
\end{equation}
using equations (\ref{dpdx}) and (\ref{Vp}). 
Note that the second term on the right-hand-side in the parentheses of equation (\ref{dpdxadi}) is small in comparison with the first term in the limit $\Gamma - 1 \ll 1$.

The internal energy therefore changes through adiabatic losses according to the relation
\begin{equation}
{dU_{adi}\over dx} = -\; m_p\;({1\over x} - {1\over 3}{d\ln \Gamma\over dx}) \;\int_0^\infty d\pp\; ({p^{\prime2}\over \gp})\Np(\pp ; x)\;,
\label{dUadidx}
\end{equation}
letting $\mu = 1$. If only adiabatic losses are important, then equation (\ref{dpdxadi}) is easily solved to give
\begin{equation}
\pp = \pp(x,x_i) = \pp_i({x_i\over x})\; [{\Gamma(x)\over \Gamma(x_i)}]^{1/3}\; ,
\label{p(x)}
\end{equation}
where $\pp_i$ is the momentum of a particle injected at radius $x_i$ when the blast wave was moving with Lorentz factor $\Gamma(x_i)$. 

The sweep-up function
\begin{equation}
Q_{su}(\pp_i,x_i) = {d\Np(\pp_i,x_i)\over d\pp_i dx_i} = 4\pi x_i^2 n_{ext}(x_i) \delta[\pp_i - P(x_i)]
\label{Qsu}
\end{equation}
is defined so that $Q_{su}(\pp_i,x_i)d\pp_i dx_i$ gives the differential number of particles with comoving-frame momentum $\pp_i$ between $\pp_i$ and $\pp_i + d\pp_i$ that are swept-up between radii $x_i$ and $x_i + dx_i$. The comoving particle distribution function at location $x$ is therefore
\begin{equation}
\Np(\pp;x) = \int_0^x dx_i |{d\pp_i\over d\pp}|\;Q_{su}(\pp,x_i)\;.
\label{Nppp}
\end{equation}
From the definition of $U$ in equation (\ref{Utot}), we obtain
\begin{equation}
U(x) = 4\pi m_p n_0 \int_0^x dx_i x_i^2 \; (\bar \gamma - 1) \; ,
\label{U(x)}
\end{equation}
where $\bar p \equiv (x_i/x)[\Gamma(x)/\Gamma(x_i)]^{1/3} P(x_i)$ and $\bar\gamma = \sqrt{\bar p^2 +1}$. Here the $\delta$-function in equation (\ref{Qsu}) is used to solve the integral over $d\pp$ in equation (\ref{Utot}), and we simplify to a uniform surrounding medium with density $n_0$. Similarly, the $dU_{adi}/dx$ term in equation (\ref{dUadidx}) is evaluated to obtain
\begin{equation}
{dU_{adi}\over dx} = -\;4\pi m_p n_0 \;({1\over x} - {1\over 3}{d\ln \Gamma\over dx}) \;\int_0^xdx_i x_i^2\; ({\bar p^2\over \bar\gamma})\;.
\label{dUadidx1}
\end{equation}

Fig.\ 1 shows a calculation of equation (\ref{dPdx}) under different assumptions for the particle energy losses. The thermal ejecta particles are assumed to be cold, so that only energy losses of the swept-up particles are considered. The parameters of the calculation are $E_0 = 10^{54}E_{54}$ ergs, $\Gamma_0 = 300\Gamma_{300}$, and $n_0 = 100n_2$ cm$^{-3}$, with $E_{54} = \Gamma_{300} = n_2 = 1$. For these parameters, the deceleration radius \citep{mr93}
\begin{equation}
x_d \equiv ( {3 E_0\over 4\pi\Gamma_0^2 m_p n_0})^{1/3} \cong 2.6\times 10^{16} ({E_{54}\over \Gamma_{300}^2 n_2})^{1/3}\;\rm{cm}\;.
\label{x_d}
\end{equation}
We note that models of afterglow spectra \citep{wg99,pk01} indicate that GRBs may take place in relatively tenuous environments with $10^{-4}$ cm$^{-3} \lesssim n_0\lesssim  \sim 10^2$ cm$^{-3}$, so that our standard parameters are on the high range of inferred values of $n_0$.  These models have neglected, however, to consider adiabatic losses of the injected electron distributions, which are important for electrons with Lorentz factors that produce synchrotron emission near the cooling break.  Moreover, the blast-wave dynamics do not take into account evolution in intermediate radiative regimes that would be important if UHECRs carry internal energy away from the system. Comparisons \citep{dbc00} of detailed numerical models with  analytic equations for afterglow emission show discrepancies by well over an order of magnitude, particularly when Compton losses are important, as is the case in many of the afterglow fits. It is not clear how these effects may change inferences about $n_0$, but the reader should keep in mind the possibility that $n_0 \sim 0.01$-1 cm$^{-3}$ is more realistic, and we will consider such densities in the subsequent discussion.

The numerical solution of equation (\ref{dPdx}), using equations (\ref{U(x)}) and (\ref{dUadidx1}) to calculate the various terms in this equation, is shown by the thick solid curve in the figure. It exhibits the behavior $P(x)\propto x^{-3/2}$ at $x \gg x_d$. This gives the $\Gamma(x)\propto x^{-3/2}$ dependence of $\Gamma$ at relativistic speeds, and the Sedov solution behavior $\beta (x)\propto x^{-3/2}$ at nonrelativistic speeds, where $P=\beta\Gamma$. For comparison, the dashed lines show the asymptotes for the shocked fluids at relativistic and nonrelativistic bulk Lorentz factors \citep{bm76}. In the relativistic case, the Lorentz factor of the shocked fluid is given by $\Gamma = (17 E/16\pi \rho_0 )^{1/2} x^{-3/2}$ when $x \gg x_d$ and $\Gamma \gg 1$, where $\rho_0$ is the rest-mass energy density of the external medium. In the nonrelativistic case, $\beta \cong 0.29(E/ \rho_0)^{1/2} x^{-3/2}$, where we assume that $\hat \gamma = 5/3$ is the adiabatic index of the external unshocked gas, and we note that the speed of the shocked fluid is $3/4$ of the speed of a nonrelativistic shock. 

The analytic solution to the equation of blast-wave evolution in the radiative regime \citep{bm76}, equation (\ref{Grad_sol}), is shown by the labeled solid curve for this case.  When $x \gg x_d$, the $P \propto x^{-3}$ behavior is recovered.  The analytic solution to the equation for blast-wave evolution with momentum-conservation, equation (\ref{Gmom}), was derived by \citet{cd99} and is shown by the second labeled solid curve in Fig.\ 1. When $x \gg x_d$, $P \propto x^{-3/2}$ when $P\gg 1$ and $P\propto x^{-3}$ when $P\ll 1$. As pointed out by \citet{hdl99}, this equation does not reproduce the Sedov solution at nonrelativistic Lorentz factors, but rather follows the momentum-conservation equation derived by Oort \citep{loz92}.

\subsection{Evolution of Comoving Particle Distribution Function and Internal Energy}

When only adiabatic losses are important, the comoving particle distribution function is given by 
\begin{equation}
\Np (\pp;x) = 4 \pi n_0\int_0^x dx_i\; x_i^2\; \tilde D \; \delta [\tilde D p^\prime - P(x_i)]\; ,
\label{Npevolve}
\end{equation}
where $\tilde D = (x/x_i) [\Gamma(x_i)/\Gamma(x)]^{1/3}$, using equations (\ref{p(x)})-(\ref{Nppp}). The internal energy $U_{adi}$, for a blast wave subject to adiabatic losses only, is given by substituting equation (\ref{Npevolve}) into equation (\ref{U}). 

The numerical results show that an adiabatic blast wave evolves according to the relation $P(x) \approx P_0$ for $x \ll x_d$, and $P(x) \approx P_0 (x/x_d)^{-3/2}$ for $x_d \ll x \ll x_{cool}$, where $x_{cool}$ is the radius when the blast wave becomes strongly radiative (see \S 2.4). Asymptotes for $\Np (\pp;x)$ and $U_{adi}$ can easily be derived using this expression for $P(x)$. Consider a blast wave that is initially relativistic so that  $P_0 \gg 1$. When $x \ll x_d$, $P(x_i) \cong P_0$, $\Gamma(x_i) \cong \Gamma(x) \cong \Gamma_0$, and $\tilde D \cong x/x_i$. One obtains
\begin{equation}
\Np (\pp;x) \cong {4 \pi n_0\over P_0^3}p^{\prime 2}\; x^3 H[\pp - P_0]\; ,\; {\rm for~}  x \ll x_d\; ,
\label{Npasymp1}
\end{equation}
and 
\begin{equation}
U_{adi} (x) \cong m_p \pi n_0 x^3 P_0 ,\; {\rm for~} x \ll x_d\; .
\label{Uasymp1}
\end{equation}
The quantity $H[a] = 1$ for $a> 1$ and $H(x) = 0$ otherwise. During the period preceeding deceleration, both the internal energy and total number of particles increases $\propto x^3$, and equation (\ref{Npasymp1}) shows that a quadratic tail of particles extending to lower energies is formed due to adiabatic losses.

During the deceleration regime $x_d \ll x \ll \Gamma_0^{2/3} x_d$ while the blast wave is still relativistic, $P(x) \cong \Gamma(x) \cong P_0 (x/x_d)^{-3/2}$, and $\tilde D = (x/x_i)^{3/2}$. One obtains 
\begin{equation}
\Np (\pp;x) \cong {4 \pi n_0 x^3\over 3}\delta[\pp - P_0 ({x\over x_d})^{3/2}]\; ,\; {\rm for~} x_d \ll x \ll \Gamma_0^{2/3} x_d\; ,
\label{Npasymp2}
\end{equation}
and 
\begin{equation}
U_{adi} (x) \cong  {4\pi m_p n_0\over 3}\; (x x_d)^{3/2} ,\; {\rm for~}  x_d \ll x \ll \Gamma_0^{2/3} x_d\; .
\label{Uasymp2}
\end{equation}
Equation (\ref{Npasymp2}) shows that during this phase, particles maintain a roughly monoenergetic distribution because adiabatic losses of particles balance the slowing down of the blast wave and the energy at which new particles are captured. At the end of the relativistic deceleration phase at $ x \cong \Gamma_0^{2/3} x_d$, the internal energy $U_{adi} \cong E_0$, that is, a large fraction of the fireball kinetic energy has been transformed into internal particle energy within the shell.

The decelerating blast wave becomes nonrelativistic when $x \gg \Gamma_0^{2/3} x_d $. In this regime, $P(x)\cong P_0 (x/x_d)^{-3/2}$,  $\Gamma(x) \cong \Gamma(x_i ) \cong 1$, and $\tilde D \cong x/x_i$.  The comoving distribution function and internal energy of particles that are swept in {\it after} the blast wave becomes nonrelativistic are
\begin{equation}
\Np (\pp;x) \cong 8 \pi n_0 \; {P_0^6 x_d^9\over p^{\prime 7} x^6}\;H[p;P_0 ({x\over x_d})^{-3/2},P_0^{2/3}({x_d\over x})] ,\; {\rm for~} x \gtrsim \Gamma_0^{2/3} x_d\; ,
\label{Npasymp3}
\end{equation}
and 
\begin{equation}
U_{adi} (x) \cong  \pi m_p n_0\; P_0^2 x_d^3 [1 - P_0^{4/3}({x_d\over x})^2 ] ,\; {\rm for~}  x \gtrsim \Gamma_0^{2/3} x_d\; ,
\label{Uasymp3}
\end{equation}
respectively.  In equation (\ref{Npasymp2}), $H[y;a,b] = 1$ for $a \leq y < b$ and $H[y;a,b] = 0$ otherwise. Comparing equations (\ref{Uasymp2}) and (\ref{Uasymp3}), one sees that the internal energy approaches a constant value.

The inset in Fig.\ 1 shows the numerical calculation of the internal energy $U_{adi}$. The above asymptotes can be seen to agree with these calculations. The difference between the adiabatic and momentum-conservation solutions can be appreciated by noting that in the absence of adiabatic losses,
\begin{equation}
U_{mom}(x) = 4\pi m_p n_0 \int_0^x dx_i x_i^2 \; [\Gamma(x_i) - 1] \; .
\label{Umom(x)}
\end{equation}
Because particles experience no adiabatic losses in the momentum-conservation solution, the internal energy will always be greater than in the adiabatic case where such losses are included. Using equation ({\ref{Umom(x)}), one sees that $U_{mom} \cong (4/3)U_{adi}$ when $x \ll x_d$, and $U_{mom} \cong 2 U_{adi}$ when $x \gg x_d$ noting that very little additional internal energy is swept into the blast wave when $x \gtrsim \Gamma_0^{2/3} x_d$.  During the episode of deceleration while the blast wave is relativistic, most of the internal energy is swept-up when $x \cong x_i$, so that $\bar p(x_i) \cong P(x)$ and $\bar\gamma \cong \Gamma(x)$ in equation (\ref{U(x)}). Thus the relativistic forms of the adiabatic and momentum-conservation solutions are similar.  Most of the internal energy that remains within the blast wave as it decelerates to nonrelativistic speeds was introduced during the relativistic phase. The greater amount of internal energy for the momentum-conservation solution means that the blast wave must travel more slowly than in the adiabatic solution in order to conserve total momentum.  

The dependences in the various limits are obtained by noting that for the momentum-conservation solution, $P \{M_0 + \int_0^x d\tilde x [dm(\tilde x)/d\tilde x ]\Gamma(\tilde x)\}\cong \beta\Gamma[M_0 + m(x) \Gamma(x)]\cong \beta \Gamma[M_0 + kx^3\Gamma ]\cong~ $const, giving the asymptotes $\Gamma \propto x^{-3/2}$ when $\Gamma_0 \gg \Gamma\gg 1$ and $\beta \propto x^{-3}$ when $\Gamma - 1 \ll 1$. For the radiative solution, $P \{M_0 + \int_0^x d\tilde x [dm(\tilde x)/d\tilde x ]\}\cong \beta\Gamma[M_0 + m(x) ]\cong \beta \Gamma[M_0 + kx^3 ]\cong$ const, giving the asymptotes $\Gamma \propto x^{-3}$ when $\Gamma_0 \gg \Gamma\gg 1$ and $\beta \propto x^{-3}$ when $\Gamma - 1 \ll 1$. The limits for the adiabatic solution would seem to be obtained through total energy conservation from the expression $\Gamma\{M_0 + $ $\int_0^x d\tilde x \Gamma(\tilde x) [dm(\tilde x)/d\tilde x ]\} \cong \Gamma[M_0 + \Gamma m(x)] \cong \Gamma[M_0 + kx^3(\Gamma -1 )]\cong~ $const. This gives the correct relativistic asymptote $\Gamma \propto x^{-3/2}$ when $\Gamma_0 \gg \Gamma\gg 1$, but implies that  $\beta \propto x^{-3}$ when $\Gamma - 1 \ll 1$. As is apparent from  the numerical results, the change in internal energy due to adiabatic losses becomes important in the nonrelativistic regime so that this estimate is not valid there.

\subsection{Nonrelativistic Limit of Equation for Blast-Wave Evolution}

The evolution of the speed of a nonrelativistic blast wave is described by the equation
\begin{equation}
-\;{d(M\beta)\over dt} = {1\over \beta}\;{dU_{adi}\over dt} + U{d\beta \over dt}\;,
\label{dMbdt}
\end{equation}
which is obtained by letting $P\rightarrow \beta$, $\Gamma\rightarrow 1$, and $dx \rightarrow \beta c dt$ in equation (\ref{dPdx}). Because $dU_{adi}/dt \simeq (\beta c)^{-1}U/x$ and $\beta \ll 1$, this term  dominates the $Ud\beta/dt$ term in equation (\ref{dMbdt}). When consideration of the external medium pressure on blast-wave evolution is taken into account, we recover the equation of \citet{cw75}. This was used to establish the important result that if SNe are the sources of cosmic rays below the knee of the CR spectrum, then CR acceleration must persist during the late phases of SNR evolution in order to overcome adiabatic energy losses within the expanding remnant.

The dotted curve in Fig.\ 1 shows the dependence 
\begin{equation}
P(x) = {P_0\over \sqrt{1+(x/x_d)^3}}\cong 
 \cases{P_0 \; ,& for $x\ll x_d$ \cr\cr
        \beta_0 ({x\over \el})^{-3/2}\; , & for $x_d \ll x \ll x_{cool}$ \cr}\;
\label{P(x)}
\end{equation}
suggested by the solution \citep{cd99} to the equation for momentum conservation. This provides a representation that is accurate to within 10\% of the numerical solution and the hydrodynamical asymptotes describing the behavior of an adiabatic blast wave.  The Sedov length
\begin{equation}
\el \equiv \Gamma_0^{2/3}x_d = ({3E_0\over 4\pi n_0 m_p })^{1/3} = 5.4\times 10^{18}\;({E_{54}\over n_0})^{1/3}\;{\rm ~cm} = 1.76 ({E_{54}\over n_0})^{1/3}\; {\rm pc}\;
\label{ell}
\end{equation}
\citep{snp96}. 

At nonrelativistic energies, $(M_0 +4\pi x^3 n_0 m/3) v = M_0 v_0$ from momentum conservation, where $v=v(x)$ and $v_0$ is the initial speed of the ejecta. Thus the SN ejecta begins to undergo significant deceleration when the swept-up mass is about equal to the ejecta mass, that is, when $4\pi \el^3 n_0 m = 3M_0$. For $\Gamma_0 - 1 \ll 1$, this occurs when $\el \approx x_d$, recalling that $E_0 \rightarrow M_0$ in this limit.

In the nonrelativistic limit with $P_0\ll 1$, equation (\ref{P(x)}) reduces to
\begin{equation}
P(x) \rightarrow \beta(x) \cong 
 \cases{\beta_0 = v_0/c \; ,& for $x\ll x_d \cong \el$ \cr\cr
        \beta_0({x\over \el})^{-3/2} = \; ({3E_{ke}\over 2\pi \rho_0})^{1/2} x^{-3/2}\; \; , & for $\el \ll x \ll x_{cool}$ \cr}\;.
\label{v0c}
\end{equation}
where $E_{ke}\equiv \beta_0^2 E_0/2$ is the kinetic energy of the injection event when $\beta_0\ll 1$, and $\rho_0 \equiv mn_0$ is the rest-mass energy density of the CBM. From the nonrelativistic branch in equation (\ref{v0c}), we see that 
\begin{equation}
\beta(x) = 0.10 ({E_{51}\over n_0})^{1/2}\;[x({\rm pc})]^{-3/2}\; ,
\label{betax}
\end{equation}
where $E_{51}$ is the explosion kinetic energy in units of $10^{51}$ ergs.  For comparison, \citet{loz92} gives the relation $\beta(x) = [4\times 2.02\times E_{ke}/(25\times \rho_0)]^{1/2} x^{-3/2} =0.086 (E_{51}/n_0)^{1/2} [x({\rm pc})]^{-3/2}$, obtained from the hydrodynamic self-similar Sedov solution.  When $x \gg x_{cool}$, the Sedov phase ends.  \citet{che74} and \citet{fal81} find that due to the onset of atomic and bremsstrahlung processes in the cooling plasma, $x_{cool}$(pc) $ = 19 E_{51}^{0.29} n_0^{-0.41}$ and $x_{cool}$(pc) $ = 20 E_{51}^{0.295} n_0^{-0.409}$, respectively (see also \citet{loz92}). 

\section{Particle Acceleration in GRB Blast Waves}

Particle acceleration can occur in GRB blast waves through a first-order Fermi mechanism involving internal or external shocks, and through second-order Fermi acceleration involving gyroresonant scattering of particles by magnetic turbulence in the magnetic field of the blast wave. \citet{wax95} and \citet{wb00} argue that shocks within a relativistic wind that persists over a timescale characteristic of the duration of a GRB could accelerate particles to UHECR energies. Such a blast wave must be highly radiative in order to give good efficiency for UHECR production; otherwise the bulk of the energy of the explosion is retained within the blast wave to be radiated during the afterglow phase when acceleration ceases and adiabatic losses reduce the energy of the nonthermal particles. UHECR acceleration in an internal shock model must also compete with strong photomeson losses that could limit acceleration to the highest energies.

An external shock acceleration model for UHECR acceleration of particles accelerated from nonrelativistic energies has been shown to be infeasible \citep{gal99}. Following the first shock crossing, subsequent energy gains are modest because cosmic rays that diffuse in the surrounding medium are captured before pitch angle scattering changes the direction of the CR with respect to the shock normal by more than $\sim 1/\Gamma$.

We therefore consider stochastic acceleration within the GRB blast wave. Wave turbulence in the magnetic field of the blast wave will be generated through the process of capturing electrons, protons, and charged dust \citep{ps00,sd00}. Rayleigh-Taylor instabilities in the expanding fluid and density irregularities in the surrounding medium will also introduce magnetohydrodynamic turbulence on sizes corresponding to the blast-wave width \citep{mrp94,rv87}. This energy will cascade to MHD turbulence on smaller size scales. Although the wave generation and cascading process is inherently dynamic, we approximate the turbulence spectrum by a power-law in wavenumber space in order to provide a simplified treatment of the process.

\subsection{Maximum Energy of Particles Retained in GRB Blast Waves}

Allowed sites of UHECR acceleration must satisfy the \citet{hil84} condition, which essentially requires that the particle Larmor radius $r_{\rm L} $ be less than the size scale $D$ of the system. For GRB blast waves, $r_{\rm L} =(m_p/eB)\pp (A/Z)$, where $B$ is the mean magnetic field, $Am_p$ is the ionic mass and $Ze$ its charge, and $D = \Dp$. When generalized for bulk relativistic motion of the acceleration region (e.g., \citet{wax95,nma95,rm98}), the maximum measured particle energy that can be accelerated in a GRB blast wave is $E_{max}$(ergs) $\cong \Gamma e Z B \Dp$.

The value of $B$ is conventionally assigned in terms of a magnetic field parameter $e_B$ that gives the magnetic field energy density in terms of the energy density of the downstream shocked fluid.   Thus 
\begin{equation}
B({\rm Gauss}) = (32\pi \mu n_0 e_B m_p)^{1/2}\sqrt{\Gamma(\Gamma-1)} = 0.39 (e_B \mu n_0)^{1/2} P \sqrt{\Gamma/(1+\Gamma)} \cong 0.39 (e_B \mu n_0)^{1/2} P\;,
\label{B}
\end{equation}
where the approximation is valid to within a factor of $\sqrt{2}$ from the extreme relativistic to the nonrelativistic regime. Recalling the definitions of $\Dp$ in equation (\ref{Vp}) and $x_d$ from equation (\ref{x_d}), we therefore see that $p \cong \Gamma \pp < 0.39 e (e_B \mu n_0)^{1/2} Z f_\Delta x \sqrt{\Gamma (\Gamma-1)}/Am_p$, implying  
\begin{equation}
E_{max}({\rm eV}) \cong 7.6\times 10^{20}Z ({f_\Delta\over 1/12}) (\mu n_2)^{1/6} e_B^{1/2} (E_{54}\Gamma_{300})^{1/3}({\Gamma\over \Gamma_0})({x\over x_d})\sqrt{1-\Gamma^{-1}}\; .
\label{Emax}
\end{equation}
The value of $E_{max}$ depends on $x$ and $P$, but we can see that at the deceleration timescale 
\begin{equation}
t_d \equiv {x_d\over P_0 \Gamma_0 c}\;,
\label{t_d}
\end{equation}
where $\Gamma \cong \Gamma_0$ and $x \cong x_d$, energetic GRBs can in principle accelerate protons and ions to $\gg 10^{20}$ eV. The weak dependence of $E_{max}$ on $n_0$ shows that UHECR acceleration is still possible for $n_0 \ll 1$ cm$^{-3}$, provided that $e_B$ approaches its equipartition value \citep{wax95}. Equation (\ref{Emax}) agrees with the expression derived by \citet{vie98a,vie95}. 

The Hillas condition when applied to electrons with $Z =1$ also yields equation (\ref{Emax}). Once ions begin escaping from the acceleration region, however, electrostatic forces may develop to modify the lepton escape rate.  Moreover, synchrotron radiation losses are much more important for leptons, as we see quantitatively in Section 3.2.6, and these losses will strongly impede lepton acceleration to energies set by this condition.

The relationship between $x$ and observer time $t \equiv t_d\tau$ is obtained by noting that $dx \cong P\Gamma c dt$. Assuming an adiabatic blast wave, we can use equation (\ref{P(x)}) to obtain
\begin{equation}
{x\over x_d} \cong 
 \cases{\tau \; ,& for $\tau \ll 1$ \cr
	(4\tau)^{1/4} \; , & for $1 \ll \tau \ll \Gamma_0^{8/3}$ \cr
        (5\tau/2\Gamma_0)^{2/5}\; , & for $\tau\gg \Gamma_0^{8/3},$ \cr}
\label{x/xd}
\end{equation}
from which it follows that 
\begin{equation}
{P\over P_0} \cong 
 \cases{1 \; ,& for $\tau \ll 1$ \cr
	(4\tau)^{-3/8} \; , & for $1 \ll \tau \ll \Gamma_0^{8/3}$ \cr
        (5\tau/2\Gamma_0)^{-3/5}\; , & for $\tau\gg \Gamma_0^{8/3}$ \cr}
\label{P/P0}
\end{equation}
and 
\begin{equation}
{\Gamma\over \Gamma_0} \cong 
 \cases{1 \; ,& for $\tau \ll 1$ \cr
	(4\tau)^{-3/8} \; , & for $1 \ll \tau \ll \Gamma_0^{8/3}$ \cr
        \Gamma_0^{-1}[1+ {P_0^2\over 2}\; ({5\tau\over 2\Gamma_0})^{-6/5}]\; , & for $\tau\gg \Gamma_0^{8/3}.$ \cr}
\label{G/G0}
\end{equation}
Expressions (\ref{x/xd})-(\ref{G/G0}) apply to initially relativistic or nonrelativistic blast waves, but in the latter case the middle branches do not obtain insofar as $\Gamma_0 \approx 1$. The late-time ($\tau \gg \gamma_0^{8/3}$) behavior of the nonrelativistic expressions persists until the blast wave becomes highly radiative. In the nonrelativistic regime, the blast wave radius depends upon observer time according to the relation $ x = (75 E_0 c^2/ 16\pi n_0 m_p)^{1/5} t^{2/5} = 0.32 (E_{51}/n_0)^{0.2} t^{0.4}({\rm yr})$ pc, in good agreement with the Sedov-Taylor self-similar solution \citep{loz92}. 

Equations (\ref{Emax}-\ref{G/G0}) show that UHECR production in GRB blast waves is in accord with the Hillas limit throughout the prompt and afterglow phase of a GRB for suitably energetic GRBs when $e_B \sim 1$. The $\sim \tau^{-1/8}$ dependence during the afterglow phase implies a reduction in $E_{max}$ by about one order of magnitude during the relativistic phase of a GRB blast wave, assuming that all other parameters remain constant. During the nonrelativistic phase, this dependence steepens to $E_{max}\propto \tau^{-1/5}$.  Insofar as the bulk of the total energy liberated by fireball transients and GRBs is emitted by the relatively infrequent, very energetic explosions with $E_{54}\gg 0.1$ \citep{der00}, GRB blast waves satisfy the Hillas condition and are viable acceleration sites for UHECR production, provided that $e_B$ approaches unity (see below).  

\subsection{Stochastic Particle Acceleration}

Our treatment of stochastic gyroresonant particle acceleration is very simplified. We make use of systematic energy gain rates and diffusive escape timescales based upon expressions derived in the quasilinear regime \citep{mel74,mr89,bls91}, and  consider only parallel-propagating modes that resonate with particles through the Doppler or anomalous Doppler resonance. The most important modes for protons and ions are shear Alfv\'en waves, including ion-cyclotron waves, whereas electrons will gyroresonate with fast mode waves, including whistlers. We assume that the spectrum of parallel waves can be described by the function $w(k)\propto k^{-q}$, where $w(k)dk$ is the differential energy density in waves with wave numbers between $k$ and $k+dk$  and $q$ is the spectral index of the turbulence. For a Kolmogorov spectrum of turbulence, $q = 5/3$. Assuming symmetry between the direction of wave propagation so that $w(k) = w(-k)$, the ratio of energy density in waves to the magnetic-field energy density $u_B = B^2/8\pi$ is given by $\zeta \equiv 2\int_{k_min}^\infty dk \; w(k)/u_B$. The smallest wave number $k_{min}$ is related to the largest size scale on which turbulence is generated. For a blast wave, we therefore assign $k_{min} \approx 1/\Delta^{\prime}$. 

In the quasilinear regime, the systematic energy-gain rate for stochastic particle acceleration implied by the pitch-angle-averaged momentum diffusion coefficient with shear Alfv\'en turbulence is given by
\begin{equation}
\langle {d\gp\over dt}\rangle_{sto} = {\pi \over 2} ({q-1\over q})\beta_A^2 \zeta (c k_{min}) (r_L^0 k_{min})^{q-2} p^{\prime q-1}\;
\label{dgdtsto}
\end{equation}
\citep{mel74,mr95,dml96}, where $c\beta_A = B/\sqrt{4\pi n^\prime m_p}$ is the Alfv\'en speed and $r_L^0 \equiv (m_p/eB)(A/Z)$ is the nonrelativistic Larmor radius of the particle, so that $r_{\rm L} = r^0_{\rm L} \pp$. Expression (\ref{dgdtsto}) also holds for gyroresonance with fast-mode waves away from the whistler branch. The magnetic-field prescription (\ref{B}) implies that $\beta_{\rm A} \simeq 2^{3/2} e_B^{1/2} P$. In order to accelerate UHECRs in GRB blast waves [eq.(\ref{Emax})], we require that $e_B \approx 1$, implying that $\beta_A\gg 1$. We interpret this unphysical result to indicate that the Alfv\'en velocity approaches $c$ whenever $P \gtrsim 1$ and $e_B \sim 1$. We also assume that $\zeta \gtrsim 0.1$, so that the magnetic field approaches the fully turbulent regime. When these conditions hold, stochastic particle acceleration is very rapid because the scattering centers are traveling at speeds approaching $c$, and the scattering rate is large.  This is contrary to the usual situation found in the interstellar medium, where $\beta_A\sim 10^{-4}$ and $\zeta$ may be $\ll 1$. 

These conditions violate, however, the quasilinear assumptions used to derive the stochastic transport coefficients leading  to equation (\ref{dgdtsto}). We nevertheless assume that this expression still provides the correct functional dependence on $\pp$ and $q$, so that 
\begin{equation}
\dot \pp_{sto} = K_p (c/\Dp)(r_L^0/\Dp)^{q-2} p^{\prime q-1}\;,
\label{pp_sto}
\end{equation}
where $K_p \cong \pi (q-1) \gamma_A^2\beta_A^2 \zeta /3q$ generalizes equation (\ref{dgdtsto}) for gyroresonant interactions with relativistic plasma waves. The kinematic factor $\gamma_A^2 = 1/(1-\beta_A^2)$ in $K_p$ allows for the possibility that the fractional change in momentum during the pitch angle isotropization time scale can exceed unity for relativistic Alfv\'en waves.  Consequently
\begin{equation}
({d\pp\over dx})_{sto} = {\dot \pp_{sto}\over Pc} = {K_p\over P r_L^0}\;({r^0_L \pp\over \Dp})^{q-1}\;.
\label{dpdxsto}
\end{equation}

Equation (\ref{pp_sto}) implies an acceleration time scale 
\begin{equation}
t_{acc} \cong | {\dot \pp_{sto}\over \pp }|^{-1} = {r_{\rm L}\over K_p c}\; ({r_{\rm L}\over \Dp})^{1-q} \gtrsim {r_{\rm L}\over K_p c}\; ,
\label{t_acc}
\end{equation}
where the inequality holds  because of the condition $r_{\rm L} \lesssim \Dp$. When $K_p = (2\pi)^{-1}$, we recover the form that \citet{rm98} use to gives the minimum acceleration time scale resulting from Fermi processes. Our subsequent comparisons agree with their conclusions when $K_p \leq (2\pi)^{-1}$, in which case $\Gamma\gg 10^2$, $n_0\gg 1$ cm$^{-3}$ and $e_B\sim 1$ are required to accelerate UHECRs. Stochastic gyroresonant acceleration with relativistic Alfv\'en waves can in principle permit $K_p$ to be $ \gg 1$, in which case such extreme values of $\Gamma$, $n_0$, and $e_B$ are not required. A detailed study of the resonant interactions of particles with relativistic plasma waves in a fully turbulent plasma is required to determine the range of possible values of $K_p$, and is beyond the scope of the present study. Here we assume that $K_p \sim 1$ in GRB blast waves.

Particles will diffuse via pitch-angle scattering and subsequently escape from the blast-wave shell. The diffusive escape timescale along the direction of the large scale magnetic field is given in the quasilinear regime by 
\begin{equation}
\tp_{esc} \approx \max[\tp_{dyn}, {\pi \over 8}(q-1)(2-q)(4-q) \tp_{dyn} \zeta (r_L^0 k_{min})^{q-2} p^{\prime q-2}]
\label{tesc}
\end{equation}
\citep{bar79,sm92,sch89,dml96}, where $\tp_{dyn} \approx \Dp/c$ represents the transit timescale for straight-line escape from the blast wave. Because much of the application in this paper falls outside the quasilinear regime, we again assume that the quasilinear proportionality holds and write
\begin{equation}
\tp_{esc} \approx \tp_{dyn}\max[1, K_t ({r_L^0  \pp\over \Dp})^{q-2}]\;.
\label{tescsto}
\end{equation}
The values that $K_t$ can assume are poorly known because $K_t$ depends on the magnetic field direction in the blast wave. If the magnetic field is primarily radial, then $K_t$ could be $\ll 1$, depending on the density of scatterers and the parallel diffusion coefficient. If there is a significant transverse magnetic field, then diffusive escape from the blast wave will be inhibited so that $K_t \gg 1$.  

\subsubsection{Comparison between Available Time and Stochastic Gain Time Scales}

The highest energy a particle can reach is limited by the time available for particle acceleration. This maximum energy can be determined by solving the equation $d\pp/dx = (d\pp/dx)_{sto}$, using equation (\ref{dpdxsto}) and noting that $r_{\rm L}^0 \Gamma$ is independent of $x$ for $\Gamma \gg 1$. Considering only proton $(A/Z = 1)$ acceleration in the relativistic regime $\Gamma\gg 1$, we find that
\begin{equation}
p_{max,av}^{\prime 2-q} \cong p_i^{\prime 2-q} + { K_p (8.0\times 10^6)^{q-2} (e_B\mu n_0)^{(2-q)/2}\over f_\Delta^{q-1}}\;(x^{2-q} -x_i^{2-q})\;,
\label{ppmax}
\end{equation}
where $\pp_i$ and $x_i$ refer to the injection momenta and location. For a Kolmogorov spectrum with $q = 5/3$, $\pp_{max,av} \cong 1.8\times 10^{-5} K_p^3 \sqrt{e_B \mu n_0}x$. For standard parameters with $x_d = 2.6\times 10^{16}$ cm and $\Gamma_0 = 300$, $p_{max,av} \cong \Gamma \pp_{max,av} \cong 1.4\times 10^{15} K_p^3 \sqrt{e_B}$ at $x = x_d$.  Thus the available time constraint does not prevent acceleration to ultra-high energies unless $K_p \ll 1$. 

\subsubsection{Comparison between Adiabatic Loss and Stochastic Gain Rates}

Particles will be stochastically accelerated until adiabatic losses prevent further acceleration. This energy is given by the condition $(d\pp/dx)_{sto} < |(d\pp/dx)_{adi}| \cong \pp/x$, where the last expression follows from equation (\ref{dpdx}), noting that $|(d\pp/dx)_{adi}| = (1+g/3)/x$ for $\Gamma(x) \propto x^{-g}$. Even for a radiative blast wave with $g = 3$, the expression $-\pp/x$ for the adiabatic energy-loss rate is valid to within a factor of 2. This relation also roughly compares the timescale for a particle to be accelerated to some value of $\pp$ against the available time. One finds that the maximum momentum that can be reached before acceleration is halted by adiabatic losses is  
\begin{equation}
p_{max,adi} = \Gamma \pp_{max,adi} = (12 K_p) ^{1/(2-q)} \; ({\Gamma \Dp\over r_{\rm L}^0})\;.
\label{pmaxadi}
\end{equation}
Recalling the Hillas condition $p < \Gamma \Dp/r_{\rm L}^0$, we therefore see that the maximum energy that a particle can attain, subject to adiabatic losses, is
\begin{equation}
E_{max,adi} = (12 K_p) ^{1/(2-q)}E_{max}\;,
\label{Emaxadi}
\end{equation}
where we have set $f_\Delta = 1/12$. Thus we see that particles can be accelerated to energies given by equation (\ref{Emax}) if $K_p \gtrsim 1/12$. It is also interesting to note that if $q = 2$, adiabatic losses will never dominate stochastic particle acceleration when $K_p \gtrsim 1/12$, and will always dominate acceleration if $K_p \lesssim 1/12$. 

Equation (\ref{Emaxadi}) is in agreement with the results of \citet{rm98} when allowance is made for the different approaches. Here we explicitly account for the spectrum of turbulence and determine $\Dp$ from hydrodynamic considerations. \cite{rm98} determine the size scale of the system from temporal variability, and define the adiabatic energy loss time scale in terms of $|\dot B/B|$ which, if $B$ is proportion to some power of $x$, yields equation (\ref{dpdxadi}) within a factor of order unity. Although our results are similar, our conclusions differ because, as noted above, we assume that $K_p$ can be $> (2\pi)^{-1}$.

\subsubsection{Comparison between Diffusive Escape and Comoving Timescales}

When the timescale for a particle with a given momentum to escape diffusively from the blast wave is shorter than the comoving timescale, then acceleration to larger momenta is no longer possible. This condition is  $K_t (\Dp/c)(r_L^0  \pp/ \Dp)^{q-2} < \tp \cong x/(cP)$, implying that $p_{max,esc}=(\Gamma\Dp/r^0_{\rm L})(K_t f_\Delta)^{1/(2-q)}$. Consequently 
\begin{equation}
E_{max,esc} \geq (K_t/12) ^{1/(2-q)}E_{max}\;.
\label{Emaxesc}
\end{equation}
The inequality applies when the transit timescale exceeds the diffusive escape timescale (eq.\ [\ref{tescsto}]). Equation (\ref{Emaxesc}) shows that if $K_t \gtrsim f_\Delta^{-1}\cong 12$, as could occur for large transverse magnetic fields in the blast wave, then particles will not diffuse from the blast wave during the available time before being accelerated to energies limited by the Hillas condition.

\subsubsection{Comparison between Stochastic Energy Gain and Diffusive Escape Timescales}

Acceleration to a given energy will be limited by the diffusive escape from the blast wave. The inverse of the timescale for stochastic acceleration is $t^{-1}_{sto}= K_p (c/\Dp) (r_{\rm L}^0 \pp/\Dp)^{q-2}$. Comparing with the diffusive escape timescale (\ref{tescsto}) gives $p < (\Gamma\Dp/r^0_{\rm L})(K_p K_t)^{1/(4-2q)}$, or
\begin{equation}
E_{max,acc/esc} = (K_p K_t) ^{1/(4-2q)}E_{max}\;.
\label{Emaxacc/esc}
\end{equation}
Thus we see that particles can be accelerated through stochastic gyroresonance   to ultra-high energies if $K_pK_t \gtrsim 1$, limited only by the condition that the Larmor radius is smaller than the blast wave width, if $K_p \gtrsim f_\Delta$ and $K_t \gtrsim 1/f_\Delta$. When $q = 2$, particles can either be accelerated to arbitrarily high energies if $K_pK_t \gtrsim 1$, or will not be accelerated if $K_pK_t \lesssim 1$.

\subsubsection{Comparison between Stochastic Energy Gain and Synchrotron Radiation Losses}

The synchrotron loss rate for particles with randomly oriented pitch angles in a magnetic field with mean strength $B$ is given by $-\dot\gp_{syn} = 4c\sigma_{\rm T} Z^4 (B^2/8\pi m_e c^2) p^{\prime 2}/[3A^3 (m_p/m_e)^3]$.  Hence
\begin{equation}
-\dot\gp_{syn} = {16\over 3}\; {\mu Z^4\over A^3}\; {c\sigma_{\rm T} e_B \mu n_0 \Gamma(\Gamma - 1)\over (m_p/m_e)^2}\;p^{\prime 2} \cong -\dot \pp_{syn}
\label{gdotsyn}
\end{equation}
Equating this rate to the stochastic energy gain rate (\ref{pp_sto}) gives the maximum particle momentum achievable due to the competition with synchrotron losses. The result is
\begin{equation}
p_{max,syn} \cong \Gamma [ {K_p\over \Dp}\; ({r_{\rm L}^0\over \Dp})^{q-2}\; {3A^3\over 16\mu Z^4}\; {(m_p/m_e)^2\over \sigma_{\rm T} e_B \mu n_0 \Gamma (\Gamma - 1)}]^{1/(3-q)}\;.
\label{pmaxsto/syn}
\end{equation}

Figs.\ 2a-d show calculations of $E_{max}$ for protons implied by the Hillas condition, equation (\ref{Emax}), compared with the maximum energy  $E_{max,syn}$(eV) $\cong 9.4\times 10^8 Ap_{max,syn} $ determined by comparing the stochastic particle acceleration energy-gain rate with the synchrotron energy-loss rate from equation (\ref{pmaxsto/syn}). Hence $A = 1$, $Z = 1$, and we let $\mu = 1$. The standard parameter set employs a Kolmogorov turbulence spectrum with $q = 5/3$, $e_B = \Gamma_{300} = n_2 = E_{54} = K_p = 1$, and $f_\Delta = 1/12$. The lower set of curves in each figure gives the Hillas condition result, and these curves are independent of the level of turbulence and particle acceleration rate set by $K_p$. The limits set by available time, adiabatic losses,  diffusive escape, and the particle acceleration rate give maximum energies proportional to $E_{max}$, as seen in Sections 3.2.1 - 3.2.4. The upper set of curves gives the maximum energy $E_{max,syn}$ determined by the competition between particle acceleration and synchrotron losses. When these latter curves are less than $E_{max}$, then the maximum possible particle energy is set by $E_{max,syn}$. 

For the standard parameter set, $x_d = 2.6\times 10^{16}$ cm and $\ell_{\rm S} = \Gamma_0^{2/3}x_d \cong 10^{18}$ cm. The dissipation of directed kinetic outflow into internal particle energy is greatest for an adiabatic blast wave over this range of distances. Thus if GRBs are the sources of UHECRs, then the values of $E_{max}$ reached between $x_d$ and $\ell_{\rm S}$ determine whether GRBs are viable sites of UHECRs. As can be seen, there are wide ranges of parameter values that permit particle acceleration to proton energies $\gtrsim 10^{20}$ eV. Because more energetic ions can have comparable Larmor radii than less energetic protons due to their larger charges, ions can reach even larger energies than protons subject to the Hillas condition. The stronger synchrotron losses for ions are not sufficient to restrict the maximum particle energy of ions to values less than that for protons, because the gyroresonant waves are more effective in accelerating the ions, which have smaller Larmor radii for a given energy. One finds that $E_{max,syn}\propto A^{(2-q)/(3-q)} (A/Z)^{(q+2)/(3-q)}$. 

Competition of Fermi acceleration with synchrotron losses has been considered previously  \citep{vie95,wax95,rm98}. Other radiative losses, in particular, photomeson production of high energy particles interacting with  synchrotron photons radiated by relativistic electrons in a GRB blast wave, can also prevent particle acceleration to ultra-high energies. \citet{wax95} and \citet{rm98} consider neutrino production losses for general Fermi acceleration processes,  \citet{wb97} consider neutrino production losses in a colliding shell model, and \citet{vie98b} considers neutrino production in a scenario where UHECRs are accelerated by first-order Fermi acceleration in relativistic GRB shocks.  However, \citet{vie98a} errs in his treatment of relativistic shock Fermi acceleration at an external shock \citep{gal99}, and so overestimates the highest energy neutrinos formed \citep{vie98a,vie98b}. Photopion processes have been treated by \citet{der00} in detail for an external shock model where UHECRs are accelerated through stochastic processes in the blast-wave shell, and shows that competition with photomeson losses does not prevent acceleration to ultra-high energies.  

Figs.\ 2a-2c show that $E_{max}$ generally increases for increasing values of $\Gamma_0$, $E_{54}$, and $n_2$, with all other parameters remaining constant. For the largest values of $E_{54}$ and $n_2$ that are considered, synchrotron losses rather than the Hillas condition determine the maximum particle energies.  Fig.\ 2d illustrates the effects of changing $K_p$, $q$, and $e_B$ on $E_{max}$. Changing $K_p$ and $q$ has no effect on $E_{max}$ but does change $E_{max,syn}$. If $K_p \lesssim 0.1$, UHECR particle acceleration becomes infeasible in GRB blast waves. The large value of $K_p$ means that the plasma is fully turbulent and has relativistic Alfv\'en speeds. Changes in $E_{max,syn}$ are only weakly dependent on the turbulence index $q$. By comparing the $q = 5/3$ and $q = 2$ results in equation (\ref{pmaxsto/syn}), one finds that the parameter dependence typically varies $\propto 1/4$ power. Because the $q = 2$ result is much less complicated, we examine the $q=2$ results for $p_{max,syn}$ from  equation (\ref{pmaxsto/syn}). When $\Gamma \gg 1$, 
\begin{equation}
p_{max,syn} \cong 4.4\times 10^{12} \; {K_p\over e_b}\; ({f_\Delta\over 1/12})^{-1}\; ({x_d\over x}) \; {A^3\over Z^4}\; ({\Gamma_{300}^2 \over \mu^2 n_2^2 E_{54}})^{1/3}\;.
\label{pmaxsyn}
\end{equation}

It is interesting to note that the standard parameters $\Gamma_{300} = E_{54} = n_2 = 1$ used here are the same as those found to give good spectral fits to gamma-ray bursts \citep{dcm00}, and these parameters also give protons with energies $\gg 10^{20}$ eV. An important difference between parameters is that $e_{B}\cong 10^{-4}$ in the spectral modeling, whereas here we let $e_B \cong 1$. The reason that $e_B$ had to be so small in the spectral modeling was to avoid forming photon spectra resulting from cooling electron distributions, which are rarely observed in GRB spectra.  But electrons will also be accelerated via gyroresonant stochastic particle processes. When particle acceleration is included, $e_B$ can be much larger because acceleration competes with radiative losses and the formation of cooling electron distributions is avoided. 

In stochastic acceleration of highly relativistic electrons, helicity effects and gyroresonance with different plasma waves are not so important as for lower energy electrons where interactions with whistler waves are important \citep{lev92}, so we can apply equation (\ref{pmaxsyn}) to determine the peak of the synchrotron spectrum radiated by electrons. Synchrotron losses rather than the Hillas condition determines the maximum energy of electrons accelerated through stochastic gyroresonance acceleration with plasma turbulence. The observed peak frequency of the $\nu F_\nu$ spectrum from accelerated electrons in a GRB source at redshift $z$ is given by
\begin{equation}
\nu_{max}\;{\rm (Hz)} = {2.8\times 10^6\over 1+z}\; B\Gamma ({p_{max}^e\over \Gamma})^2\;
\label{numax}
\end{equation}
for $q = 2$, where $p_{max}^e$ is $(m_p/m_e)^2$ smaller than $p_{max,syn}$ given by equation ({\ref{pmaxsyn}), and $A = Z =1$. Thus
\begin{equation}
E_{pk}\;{\rm (keV)} = h\nu_{max} = {75\over 1+z}\; ({x_d\over x})^2\; ({f_\Delta\over 1/12})^{-2}\; {K_p^2 \Gamma_{300}^{4/3}\over e_B^{3/2}\mu^{5/6} n_2^{5/6} E_{54}^{2/3}}\;
\label{hnumax}
\end{equation}
Equation (\ref{hnumax}) shows that $E_{pk}$ is in the range of $E_{pk}$ values observed with BATSE \citep{mal95} between $\approx 100 $ keV and $1$ MeV when $\Gamma_0 \sim 300$-3000. This effect may resolve a fine-tuning issue in the external shock scenario \citep{cd99} whereby the value of $e_B$ has to be large enough to provide good radiative efficiency, but not so large that cooling electron distributions are formed.

Multiwavelength afterglow modeling within the standard blast wave model \citep{wg99,pk01} imply that $n_0$ and $e_B$ are much smaller than the standard values used here, as already noted in Section 2.2. If these implied values are correct, then acceleration of UHECRs in GRB blast waves is very difficult during the afterglow phase, as is apparent from equation (\ref{Emax}). For example, consider the parameters derived by \citet{wg99} for GRB 990508, namely $n_0 = 0.03$ cm$^{-3}$, $E_{54} = 0.03$, and $e_B = 0.12$. Acceleration of $\gtrsim 10^{20}$ eV protons requires unreasonably large initial Lorentz factors, namely $\Gamma_0 \sim 3\times 10^{4}$.  Similarly conclusions are obtained using parameters for GRB 980703 and GRB 990123 derived by \citet{pk01}.

This may imply that the bulk of UHECR acceleration is done by GRBs during the prompt phase where the derived parameters may not hold. The inclusion of stochastic particle acceleration within the blast wave itself represents, moreover, a fundamental departure from the standard blast wave model.  Acceleration and radiation losses take place concurrently in the model under consideration, as compared with the standard assumption where particles are instantaneously accelerated and subsequently cool. Moreover, escape of UHECRs from the blast wave may drive the system into different radiative regimes that will complicate parameter estimation. An important test to check the derived value of the blast-wave magnetic field is to observe the appearance of the synchrotron-self Compton component in X-ray afterglow spectra, because its intensity is very sensitive to $e_B$ \citep{dcm00}.

\section{Solutions to Particle Continuity Equation}

We examine particle acceleration, escape, and losses within a GRB blast wave by employing a particle continuity equation\begin{equation}
{\partial N (p;t)\over \partial t} + {\partial\over \partial p}[\dot p(p,t) N(p;t)] + {N(p;t)\over t_{esc}(p,t)} =  Q(p,t )\;.
\label{pconteq}
\end{equation}
The term $\dot p$ represents the total energy-change rate of particles due to acceleration, adiabatic and radiative processes, and $t_{esc}$ is the particle escape timescale from the blast wave. Primes have been dropped to simplify the notation. The analytic solution to this equation when $\dot p = \dot p(p)< 0 $ and $t_{esc} = t_{esc}(p)$ are independent of $t$ is 
\begin{equation}
N (p;t) = {1\over |\dot p |} \int _p^\infty dp^* Q(p^*,t^*)\exp[-\int_p^{p*}\;{dp^{\prime\prime}\over t_{esc}(p^{\prime\prime})|\dot p(p^{\prime\prime})|}]
\label{N(p;t)}
\end{equation}
where $t^* = t - \int_p^{p*} dp^{\prime\prime}/|\dot p(p^{\prime\prime})|$. The solution for $\dot p > 0$ is obtained by reversing the limits on the integrals in the exponential and in the definition of $t^*$, and by integrating from 0 to $p$ in the integration over $p^*$.

When the terms $\dot p$ and $t_{esc}$ depend on both $p$ and $t$, the solution to equation (\ref{pconteq}) can be obtained semi-analytically by inspection, giving
\begin{equation}
N(p;t) = \int_0^t dt_i \; Q[p_i(p,t,t_i),t_i]\; |{dp_i\over dp}|\; \exp\{ -\int_{t_i}^t {dt^* \over t_{esc}[p_i(p,t,t^*),t^*]}\}\;,
\label{npt_solt}
\end{equation}
where $p_i = p_i(p,t,t^{\prime\prime})$ is obtained by inverting $p = p(p_i,t,t^{\prime\prime})$, which solves $dp/dt = \dot p$. Equation (\ref{npt_solt}) can be shown to reduce to equation (\ref{N(p;t)}) when $t_{esc}$ and $\dot p$ are time-independent.

For particle acceleration in a blast wave, it is more convenient to consider the evolution of the comoving distribution function with location $x$. Using the relationship $dx = Pcd\tp$ between radius and comoving time, equation (\ref{pconteq}) becomes
\begin{equation}
{\partial N (p;x)\over \partial x} + {\partial\over \partial p}\{[{\dot p_m\over Pc}+({dp\over dx})_{adi}] N(p;x)\} + {N(p;x)\over Pc t_{esc}(p,x)} =  Q(p,x) \;.
\label{pconteqx}
\end{equation}
 The term $\dot p_m$ represents the sum of the particle energy gain and loss rates, excluding adiabatic losses which are given through the term $(dp/dx)_{adi}$ from equation (\ref{dpdxadi}). Following equation (\ref{npt_solt}), the solution to equation (\ref{pconteqx}) is  
\begin{equation}
N(p;x) = 4\pi n_0\int_0^x dx_i \; C(x_i) x_i^2\;\delta[p_i(p,x,x_i)-P(x_i)]\; |{dp_i\over dp}|\; \exp\{ -\int_{x_i}^x {dx^* \over Pct_{esc}[p_i(p,x,x^*),x^*]}\}\;.
\label{npx_solt}
\end{equation}
Here we substitute equation (\ref{Qsu}) for the source function $Q$, though modulated by the factor $C(x_i)$. Equation (\ref{npx_solt}) can be solved analytically in some restricted cases, but must be solved numerically for the general case involving adiabatic losses and stochastic energy gains. When the energy lost through particle escape is a large fraction of the swept-up energy, the system departs from its adiabatic behavior.  In this case, a more general numerical approach involving inversion of a tri-diagonal matrix must be used. We describe the numerical considerations for dealing with this system, though here we only report the results for blast waves that evolve in the adiabatic regime.

\subsection{Analytic Solution}

Equation (\ref{npx_solt}) can only be solved analytically when the turbulence index $q = 2$. Depending on how the MHD turbulence develops and evolves, this case may be of primary interest. It is generally thought that wave energy is introduced on the largest length scales of the system and is transferred to smaller scales through nonlinear mode coupling \citep{zh90}, to be dissipated through particle acceleration when the particle Larmor radii are in resonance with cascading turbulence. In the Kolmogorov phenomenology, the wave energy achieves a steady-state turbulence spectrum with index $q = 5/3$.

Particles can only be accelerated when $(dp/dx)_{sto} > |(dp/dx)_{adi}|$ (Section 3.2.2), that is, when $p \gtrsim (\Delta/r_{\rm L}^0)(f_\Delta/K_p)^{1/(q-2)}$ if $q > 2$, and when $p \lesssim (\Delta/r_{\rm L}^0)(f_\Delta/K_p)^{1/(2-q)}$ if $q < 2$. If $q > 2$, most of the MHD turbulence energy is contained in small wavevector (long wavelength) turbulence that accelerates high-energy particles with large Larmor radii, and vice versa for $q < 2$.  Consequently, there will be a resonance gap that prevents particle acceleration when $q > 2$, because adiabatic losses dominate particle acceleration except for the very highest energy particles, of which there may be none if $(\Delta/r_{\rm L}^0)(f_\Delta/K_p)^{1/(q-2)} > P_0$. If the turbulence spectrum evolves to the Kolmogorov index through a progressive hardening of the turbulence spectrum, then the low-energy swept-up particles will begin to sample the evolving turbulence spectrum when $q = 2$. If $K_p \gtrsim f_\Delta$  and $q = 2$, stochastic acceleration will dominate adiabatic losses and accelerate particles to the highest energies.  The waves will be damped by transferring their energy to the particles. In this way, a quasi-steady turbulence spectrum with $q \cong 2$ may be formed. 

When $q = 2$ and $K_p \gtrsim f_\Delta$, the stochastic energy gain rate dominates the adiabatic energy-loss rate for all particle momenta. Thus $(dp/dx)_{sto} \cong K_p p /(P\Delta )$ from equation (\ref{dpdxsto}). In the limit $P \gg 1$, $(dp/dx)_{sto} \cong K_p p /(f_\Delta x)$ and, provided $K_t > 1$, $t_{esc} \cong K_t \Delta/c$ from equation (\ref{tescsto}). The particle momenta therefore evolve according to the relation $p = p_i(x/x_i)^{K_p/f_\Delta}$. Equation (\ref{npx_solt}) becomes 
\begin{equation}
N(p;x) = 4\pi n_0\int_0^x dx_i \; C(x_i) x_i^2\;\delta[p({x_i\over x})^{K_p/f_\Delta} -P(x_i)]\; ({x\over x_i})^{-(K_p + 1/K_t)/f_\Delta}\;.
\label{npx_ana}
\end{equation}
There is no high-energy cutoff to the accelerated particle momentum in this limit, because the acceleration rate $(dp/dx)_{sto} \propto 1/x$ and thus rapidly accelerates particles at the earliest times. 

When $x_i \ll x_d$, $P(x_i) \cong P_0$ and 
\begin{equation}
N(p;x) = 4\pi n_0\;C(\bar x_i) \; ({f_\Delta x^3\over K_p P_0})\; x^3 \; ({P_0\over p})^{1+ 3f_\Delta/K_p + 1/(K_pK_t)}\;,
\label{npx_ana1}
\end{equation}
where $\bar x_i = x({P_0\over p})^{f_\Delta/K_p}$. When $x_i \gg x_d$, $P(x_i) \cong P_0 (x_i/x_d)^{-3/2}$ and
\begin{equation}
N(p;x) = {4\pi n_0\;C(\tilde x_i) \; \tilde x_i^2 \; (x/\tilde x_i)^{-(K_p + 1/K_t)/f_\Delta}\over 
[K_p p (\tilde x_i/ x)^{K_p/f_\Delta}/f_\Delta \tilde x_i] + [3P_0(\tilde x_i/x)^{-3/2}/2\tilde x_i]}\;,
\label{npx_ana2}
\end{equation}
where $\tilde x_i= \ (P_0 x_d^{3/2} x^{K_p/f_\Delta}/p)^{1/(K_p/f_\Delta +3/2)}$

Fig.\ 3 shows the comoving particle distribution function that results from stochastic gyroresonant acceleration by plasma turbulence with $q = 2$ and $K_p = 0.5$. We let $K_t = 2$ in Fig.\ 3a and $K_t = 10$ in Fig.\ 3b. The other parameters are  the same as used in Figs.\ 1 and 2, namely $E_{54} = n_2 = \Gamma_{300} = e_B = 1$, and we let the modulation factor $C(x_i) = 1$. The particle distribution is multiplied by the factor $m_ec^2 p\gamma$ in order to indicate the energy contained in the accelerated particles. The comoving particle distribution is well-described by a power law spectrum, and the index depends sensitively on the assumed values of $K_p$ and $K_t$. For these parameters, particles can reach ultra-high energies before the blast wave decelerates to nonrelativistic speeds, which occurs at $x \cong 45 x_d$ for the chosen parameters. Because the escape timescale is independent of particle energy, power-law distributions of particles also escape from the blast wave. The highest energy escaping particles with large Larmor radii can freely escape to become UHECRs. In contrast, escaping particles with lower energies and smaller Larmor radii will be subject to the cosmic-ray adiabatic loss problem if their energy density exceeds the magnetic-field energy density in the local ISM.

\subsection{Numerical Solution}

To solve the continuity equation (\ref{npx_solt}) numerically, we discretized this equation using a conservative finite differencing scheme.  To this end we introduce
\begin{equation}
N_j^i = N(x_i,p_j)\; ;\;\; F_j^i = \dot p(x_i,p_j)N(x_i,p_j)\;.\;\; 
\label{Nij}
\end{equation} Equation (\ref{npx_solt}) becomes
\begin{equation}
{N_j^{i+1}-N_j^i \over \Delta x} = {F_{j+1/2}^{i+1}-F_{j-1/2}^{i+1} \over \Delta p} - {N_j^{i+1}\over t_{esc}(x_i,p_j)} + Q_j^i\; \;
\label{Nijcont}
\end{equation}
We make the identifications $N_j^{i+1} =N_i^j$ and $N_j^{i+1} =N_i^j$. This leads to larger diagonal elements in the resulting system of equations and enhances the stability of the numerical algorithm. As a result, we are led at each $x$-step to a tridiagonal system of equations for $N_j^i$, namely
\begin{equation}
a_j N_{j+1}^{i+1} + b_j N_j^{i+1} + c_j N_{j-1}^{i+1} = S_j^i\;,
\label{a_j}
\end{equation}
where $a_j = 0$, $b_j = 1 +\Delta x [ t^{-1}_{esc}(x_i, p_j) - \dot p/\Delta p]$, $c_j = \dot p\Delta x /\Delta p$, and $S_j^i = Q_j^i + N_j^i$. In our computations, the grid in $p$ consisted of $10^4$ intervals, with a linear spacing in the range $1 \leq p < 10^5$, and a logarithmic spacing in the range $10^5\leq p < 10^9$. The step size in x $\Delta x = x_{j+1} - x_j$ was determined adaptively at each step to ensure that system (\ref{a_j}) is well-behaved. The resulting $\Delta x$ varied between $10^{10}$ and $10^{12}$ cm during the simulations. 

Figs.\ 4 and 5 show the numerical simulation results for a Kolmogorov turbulence spectrum $q = 5/3$ when the blast wave reaches different locations between $0.1 x_d$ and $10 x_d$. The evolution of the blast wave is assumed to follow the adiabatic behavior described by equation (\ref{P(x)}). We use $K_p = 0.1$, and $K_t = 10$ in Fig.\ 4, and $K_p = 1$ and $K_t = 10$ in Fig.\ 5. The other parameters are the same as used in Fig.\ 4. Panel (a) in the figures show the time evolution of the comoving particle distribution, and panel (b) shows the cumulative energy distribution function of escaping particles measured by an outside observer, obtained by solving
\begin{equation}
N_{esc}(p_{esc}; x) = \int_0^\infty d\bar x \; {N_{esc}[p_{esc}/\Gamma(\bar x),\bar x]\over \Gamma(\bar x)}\;,
\label{Nesc}
\end{equation}
where $N_{esc}(p,x) = N(p;x)/Pct_{esc}(p,x)$ from equation (\ref{pconteqx}).
 
In Fig.\ 4, we let the modulation factor $C(x_i) = 1$.    The development of the internal comoving particle distribution 
is shown in Fig.\ 4a at different blast-wave locations. The particles diffuse to higher energies with time, reaching a maximum value that is about an order-of-magnitude less than the maximum momentum permitted through the available-time constraint that was estimated from equation (\ref{ppmax}). This equation does not, however, take into account adiabatic losses which appreciably retard acceleration of the highest energy particles. Fig.\ 4b shows the escaping particle distribution that would be measured in an external system. Most of the energy is carried by the very highest energy particles with energies between $10^{18}$ and $10^{20}$ eV.  

The escaping UHECRs carry more than $10^{56}$ ergs of energy, because wave damping and energy conservation have not been taken into account in the calculation.  A realistic calculation must consider these processes, as in treatments of impulsive Solar flares \citep{mil98,mr95}. Moreover, we have restricted the blast wave evolution to the adiabatic regime. The calculation of Fig.\ 4 does not violate energy conservation or the assumption of an adiabatic blast wave if we assume that $C(x_i) \lesssim 10^{-3}$, so that only a very small fraction of swept-up particles are accelerated to the highest energies. It is artificial to assume that $C(x_i)$ does not change with time location, but a self-consistent treatment of wave cascading and damping, required to obtain a correct evaluation of $C(x_i)$, is beyond the scope of this paper. 

In Fig.\ 5 we show a calculation of the evolving particle distribution with $K_p = 1$.  Here we assume that $C(x_i) = 10^{-4}$ so that the calculation is self-consistent. The effect of increasing the rate constant $K_p$ is to permit particle acceleration to  much higher energies, until adiabatic losses and escape suppress further acceleration.  Fig.\ 5b shows that a significant fraction of the energy is carried by particles with energies $\gtrsim 10^{20}$ eV. In this calculation, the cumulative escaping particle spectrum develops a spectrum with number index $\sim -3$. Larger values of $K_t$ or other parameters of the calculation, such as $\Gamma_0$, can yield even higher energy escaping particles. Thus stochastic particle acceleration in relativistic blast waves can, in principle, produce cosmic rays with energies $\gtrsim 10^{20}$ eV. 

\subsection{Cosmic Ray and Ultra-High Energy Cosmic Ray Origin}

As shown in Figs.\ 4 and 5, protons accelerated to $\gtrsim 10^{19}$ eV by a Kolmogorov MHD turbulence spectrum can diffusively escape from the blast wave during the prompt and afterglow phases of a GRB to become UHECRs. The Larmor radii of these particles are so large that they do not strongly couple with the interstellar gas and thereby avoid further adiabatic losses. This is because the energy density of UHECRs within a volume defined by their Larmor radius is smaller than the energy density of the mean magnetic field within that volume.  UHECRs consequently behave as isolated particles rather than as a relativistic fluid that does work on its surroundings.  

This is not the case for lower energy particles which, if they escaped into the ISM, would generate streaming instabilities that would strongly couple these particles with the ISM. The adiabatic loss problem would be severe for these particles. As shown by the simulations, however, only the very highest energy particles escape during the relativistic phases of the blast wave. Once the blast wave enters the nonrelativistic Sedov phase, stochastic acceleration to ultra-high energies is much weaker because the Alfv\'en speed becomes $\ll c$.  During this phase, shock Fermi acceleration becomes more efficient than stochastic acceleration. The internal particle population confined by the blast wave loses energy due to adiabatic expansion of the blast wave shell, but also begins to be energized through shock acceleration. The nonrelativistic phase of a GRB remnant evolution resembles a very energetic SNR, though differing from a conventional SNR by containing a population of high energy particles available from the earlier acceleration episode.  Consequently the timescale constraint \citep{lc83} on shock Fermi acceleration is relaxed. Cosmic rays will leave the remnant as the SNR dissipates in the ISM so that the cosmic-ray adiabatic loss problem is solved, as in the standard model, by shock acceleration taking place throughout the expansion of the SNR until the energy density of the relativistic particle fluid is small compared to the magnetic-field energy density of the ISM.

Future work must treat both first- and second-order particle acceleration in a relativistic blast wave that decelerates to nonrelativistic speeds and determine whether this scenario can satisfactorily explain the origin of the hadronic cosmic rays above the knee of the cosmic ray spectrum, as well as making a contribution to hadronic CR origin at lower energies.

\section{Summary and Conclusions}

In this paper, we have treated adiabatic losses stochastic particle acceleration in blast waves formed by the ejecta from exploding stars. Because the Alfv\'en speed approaches $c$ whenever the magnetic field approaches its equipartition value in relativistic shocks, second-order Fermi acceleration can be much more efficient than first-order Fermi acceleration in GRB blast waves.  In nonrelativistic SNR shock waves, by contrast, the first-order process dominates. Consequently, we have considered particle acceleration through stochastic gyroresonant acceleration in GRB blast waves during their relativistic phases. Even in the highly simplified approach developed here, it was necessary to treat adiabatic losses correctly in the expanding blast wave. This was dealt with in Section 2, where an equation for blast-wave evolution was derived (equation (\ref{dPdx})) that contains a term that rechannels the energy lost by adiabatic expansion of the nonthermal particles into the directed outflow of the blast wave. An expression for the energy lost through adiabatic expansion by a general particle distribution function that contains both relativistic and nonrelativistic particles was derived in Section 2.2. The evolution of the internal particle distribution due to adiabatic losses was treated in Section 2.3, and an expression describing the evolution of an adiabatic blast wave which agreed with numerical calculations, was shown to reduce to well-known results in the nonrelativistic regime in Section 2.4.

In Section 3 we showed that ultra-high energy particle production in GRB blast waves satisfies the Hillas condition, which compares the particle Larmor radius with the size scale of the system. In the case of the GRB system, the size scale is the blast-wave width.  Thus GRBs are allowed sites for UHECR production if the magnetic field approaches its equipartition value. We employed simplified forms for the energy gain rate and escape time scale due to stochastic gyroresonant interactions of particles with MHD turbulence based upon expressions derived in the quasilinear approximation. In Section 3.2, we derived the conditions under which competition between stochastic acceleration, available time, adiabatic losses, and diffusive escape permits acceleration to ultra-high energies.  The conditions under which synchrotron losses limit particle acceleration were also examined. We find that wide ranges of parameter values permit particle acceleration to ultra-high energies through stochastic gyroresonant processes in GRB blast waves. However, if parameters derived in afterglow modeling are correct, then many GRBs cannot accelerate UHECRs during the afterglow phases. In Section 4, we numerically calculated comoving and escaping particle distributions by employing a continuity equation, and found that a population of UHECRs will diffusively escape from GRB blast waves during the prompt and afterglow phases.

This study is preliminary and does not address the question of the origin of the turbulence, nor wave cascading and damping. We also assume that the blast wave is homogeneous, whereas it is in fact subject to instabilities and nonlinear effects on the shock structure from the accelerated particle distribution. Within the constraints of the problem that are rather well defined, for example, the maximum blast-wave magnetic field, the characteristic blast-wave width, the adiabatic loss rate, and the time available for acceleration, we conclude that particle acceleration to ultra-high energies is possible in GRB blast waves. This process therefore offers a possible solution to UHECR acceleration and the origin of hadronic cosmic rays above the knee of the cosmic-ray spectrum. 

\acknowledgments{We thank A. Achterberg, M. Baring, L. O'C. Drury, A. Levinson, J. A. Miller, and H. V\"olk for discussions, and the referee for constructive criticisms.  The work of C.D. is supported by the Office of Naval Research and the NASA Astrophysics Theory Program (DPR S-13756G).}


\clearpage

\begin{figure}
\vskip-1.0in
\figurenum{1}
\epsscale{0.9}
\plotone{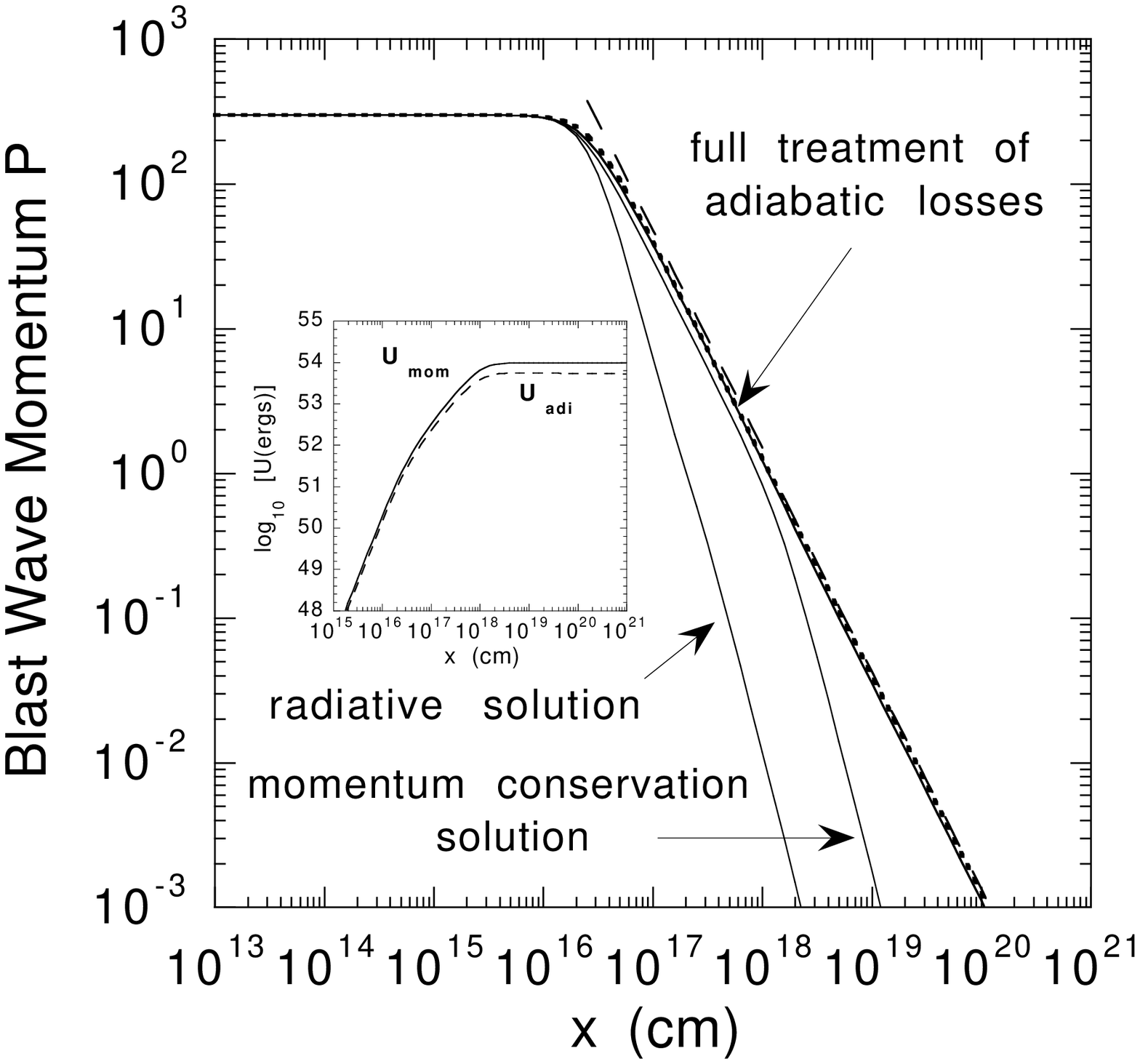}
\caption{Evolution of blast wave momentum $P$ for different assumptions for the particle energy losses. The solid curve labeled ``full treatment of adiabatic losses" is a numerical solution of equation (\ref{dPdx}) that includes adiabatic losses of swept-up particle energy that is transformed into the directed kinetic energy of the outflow. The curve labeled ``momentum conservation solution" neglects adiabatic energy losses of the swept-up particles, and the curve labeled ``radiative solution" assumes that the internal energy is promptly radiated.  Dashed lines are hydrodynamic approximations at relativistic and nonrelativistic momenta, and the dotted curve is equation (\ref{P(x)}).   The inset shows the evolution of the internal energy $U_{mom}$ and $U_{adi}$ for the momentum-conservation and adiabatic solutions, respectively.
}
\end{figure}

\begin{figure}
\figurenum{2}
\epsscale{1.0}
\plottwo{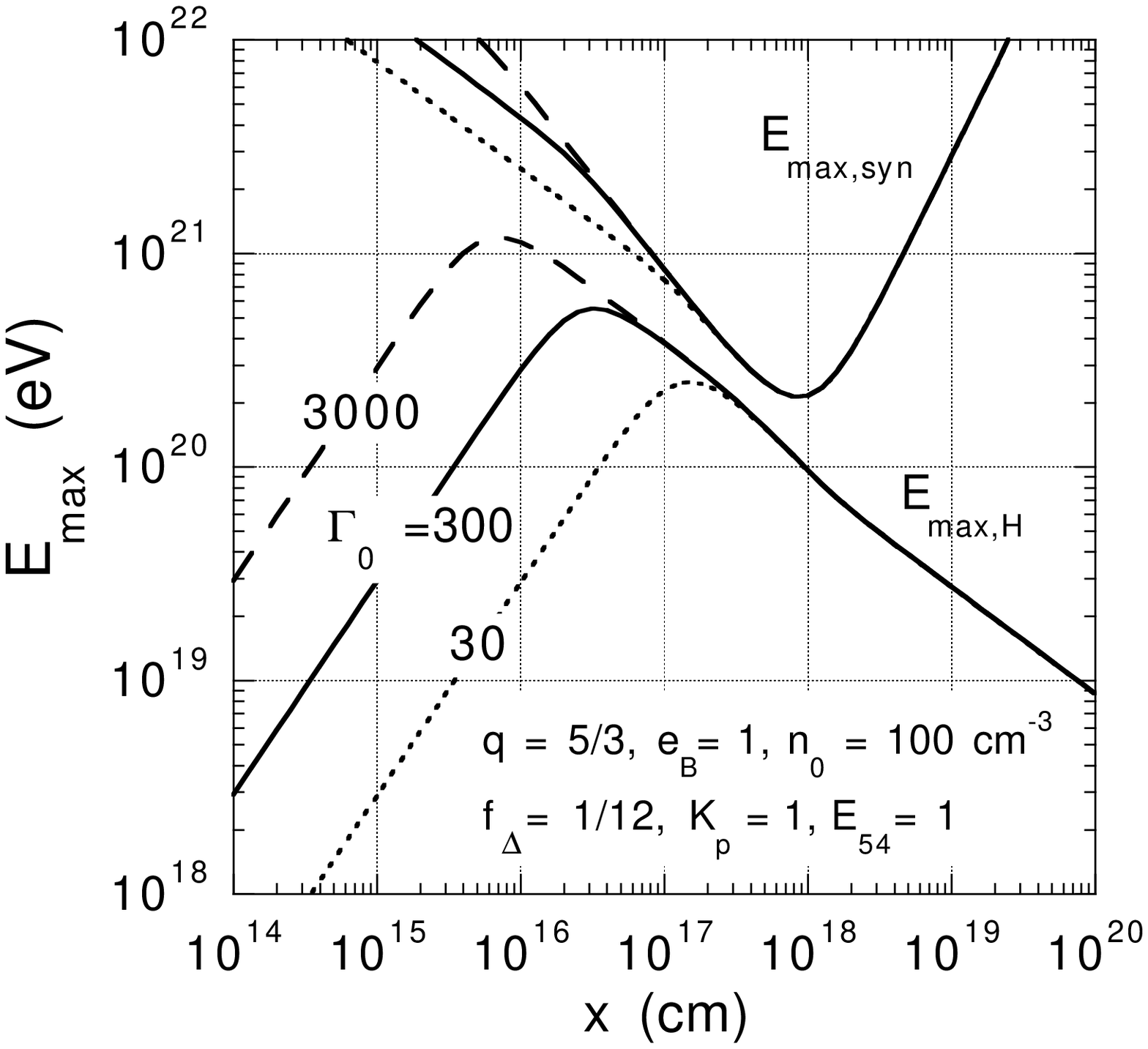}{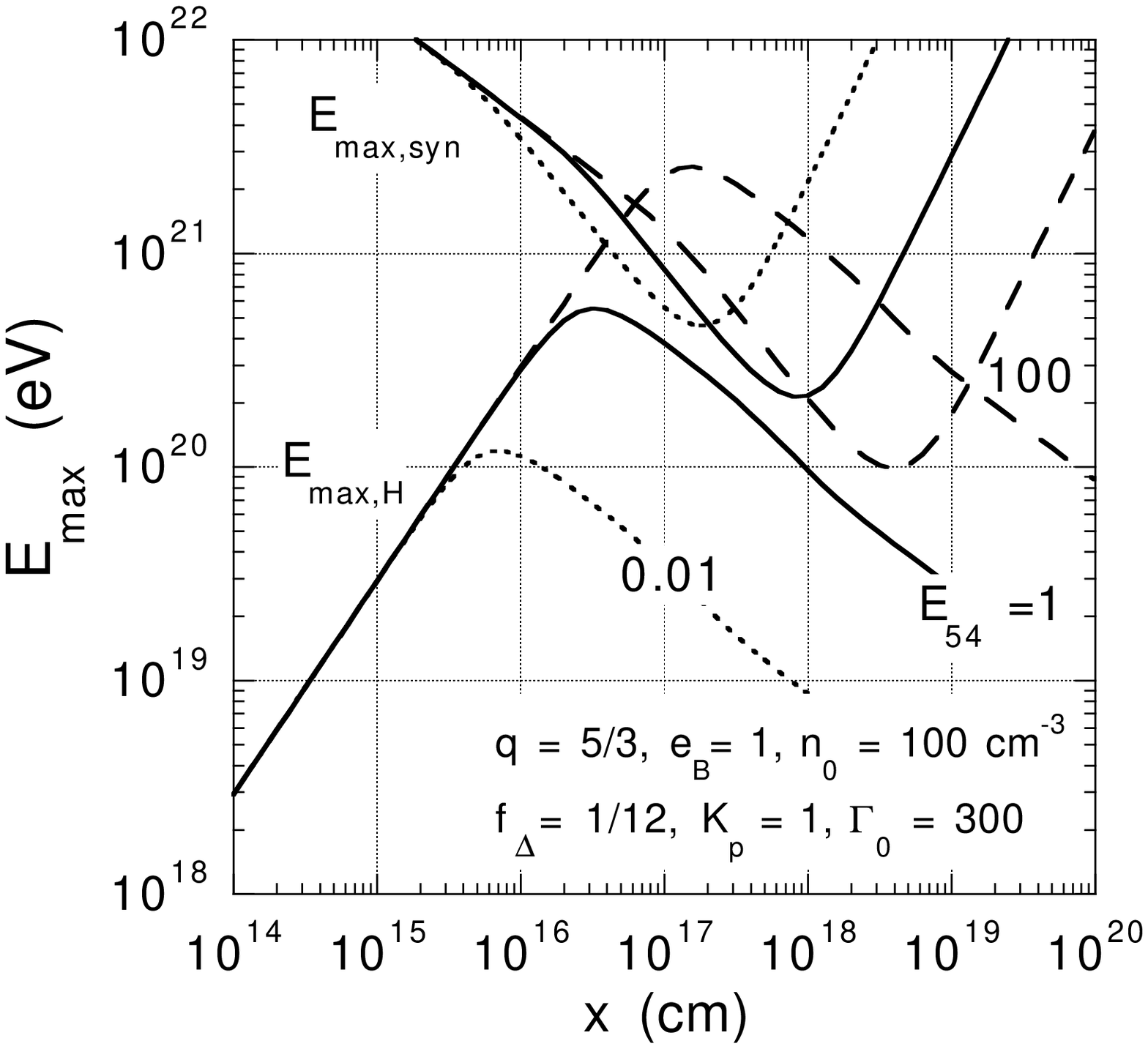}
\caption{Maximum proton energies $E_{max,{\rm H}}$ as a function of blast wave location $x$ implied by the Hillas condition, equation (\ref{Emax}), obtained by comparing the Larmor radius and blast wave width, and $E_{max,syn}$ from equation (\ref{pmaxsto/syn}), obtained by comparing the stochastic acceleration rate with the synchrotron cooling rate. Standard parameters are shown in the figure legend. Changes in $E_{max,{\rm H}}$ and $E_{max,syn}$ for different values of $\Gamma_0$, $E_{54}$, and $n_0$ are shown in (a), (b), and (c), respectively. Changes in $E_{max,syn}$ for different turbulence spectral indices $q$ and acceleration rates $K_p$, and $e_B$ are shown in (d), as well as changes in $E_{max,{\rm H}}$ and $E_{max,syn}$ due to changing $e_B$ values.
}
\end{figure}

\begin{figure}
\figurenum{2}
\epsscale{1.0}
\plottwo{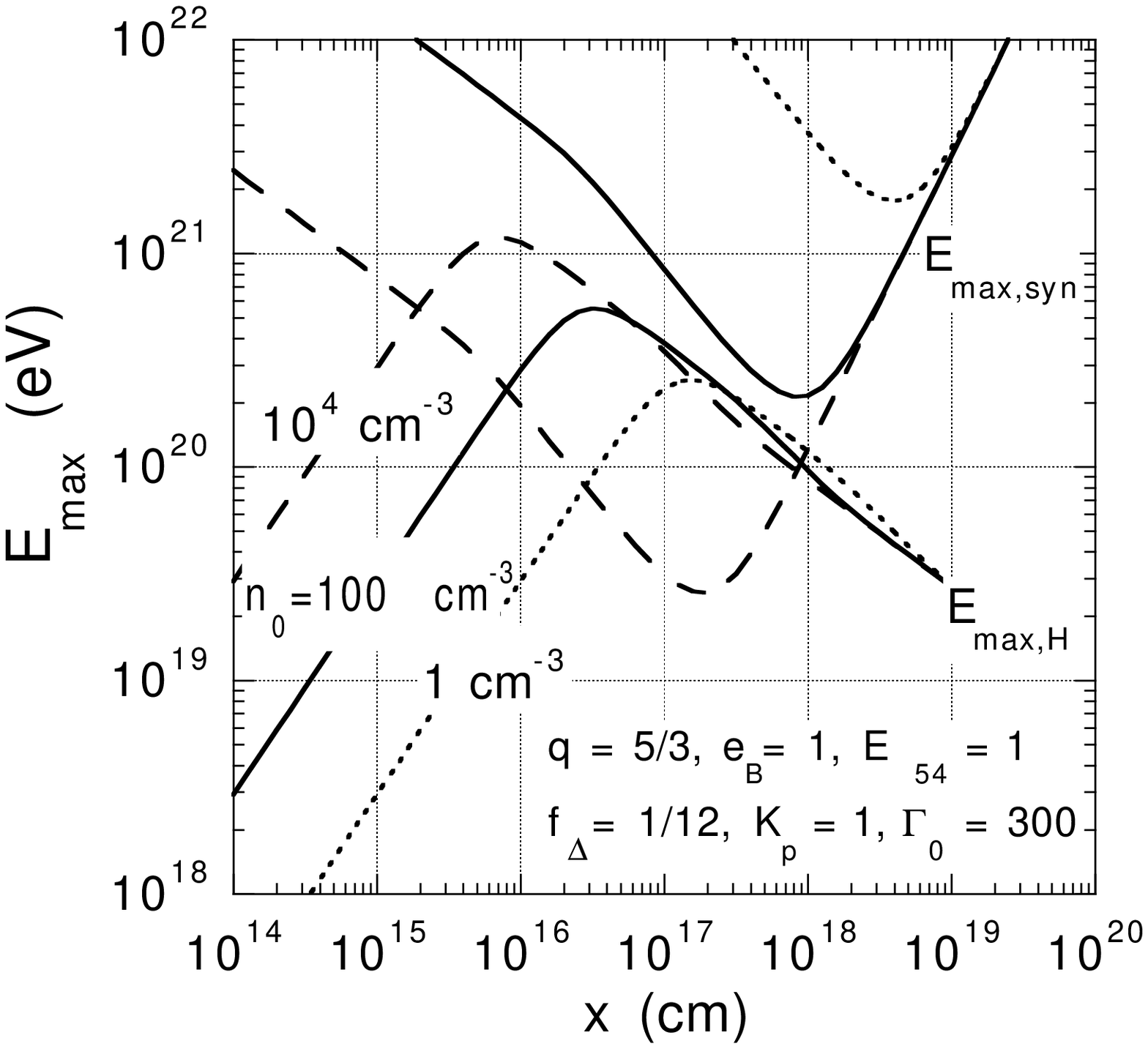}{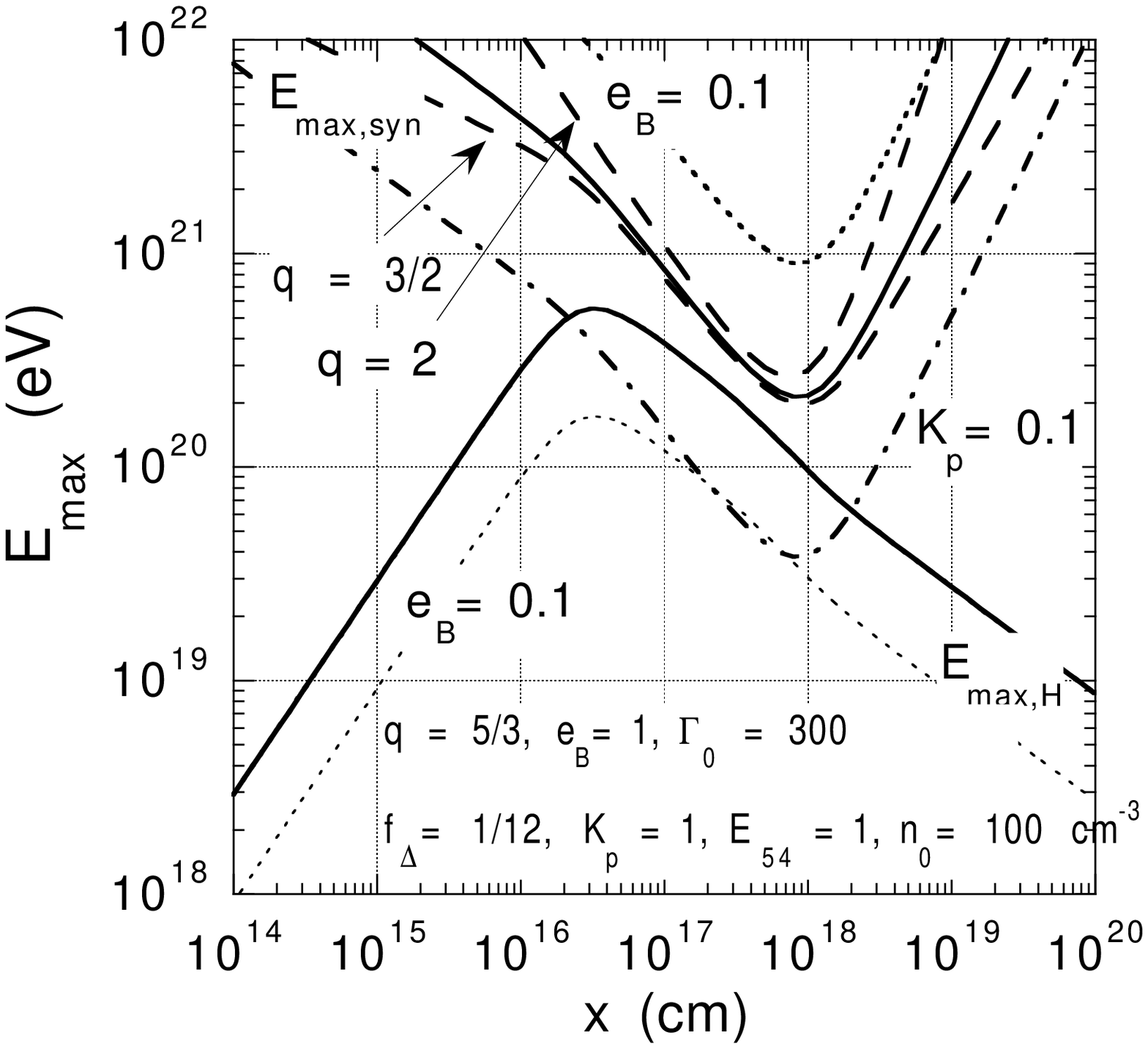}
\caption{Fig.2c (left), Fig. 2d (right)
}
\end{figure}

\begin{figure}
\figurenum{3}
\epsscale{1.0}
\plottwo{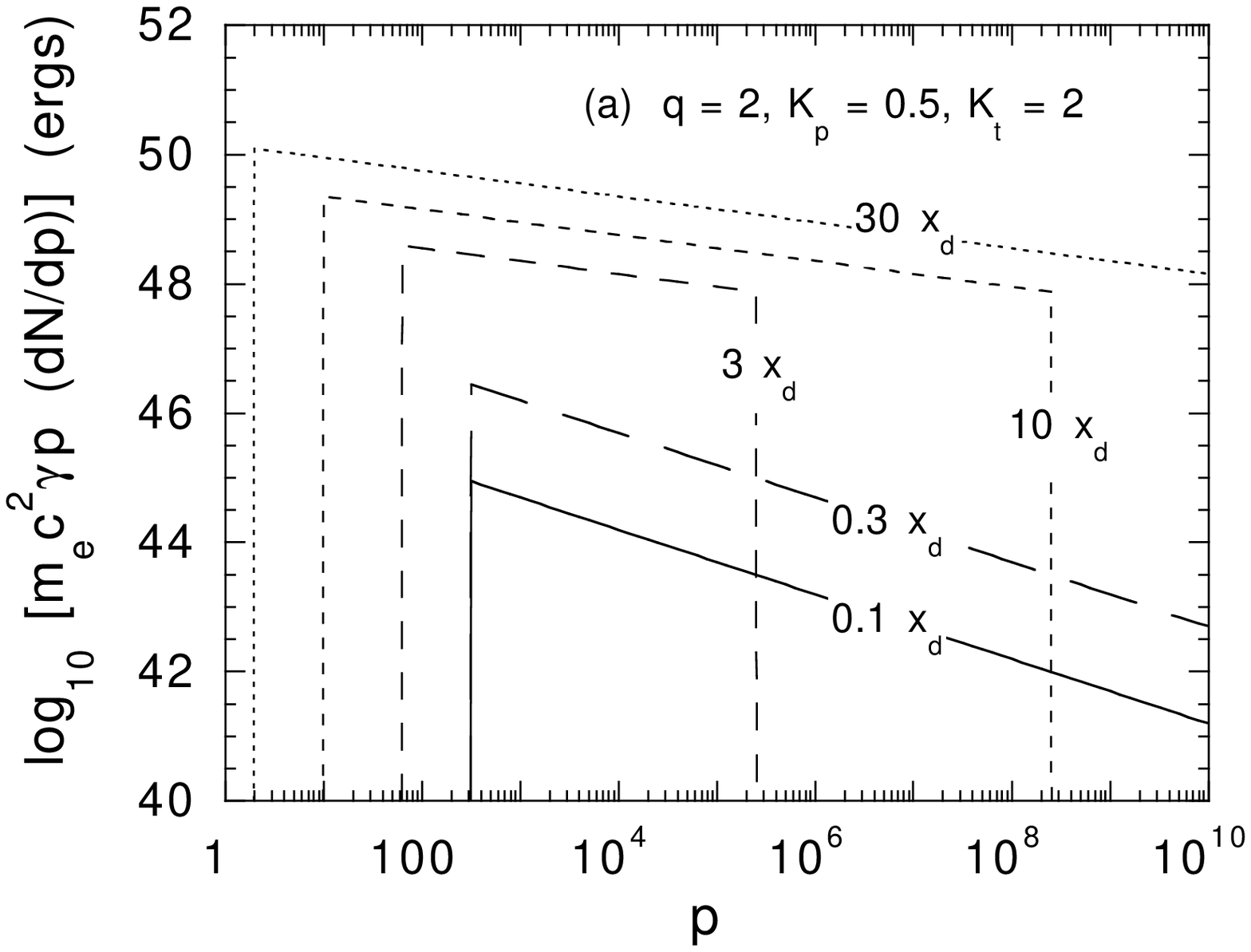}{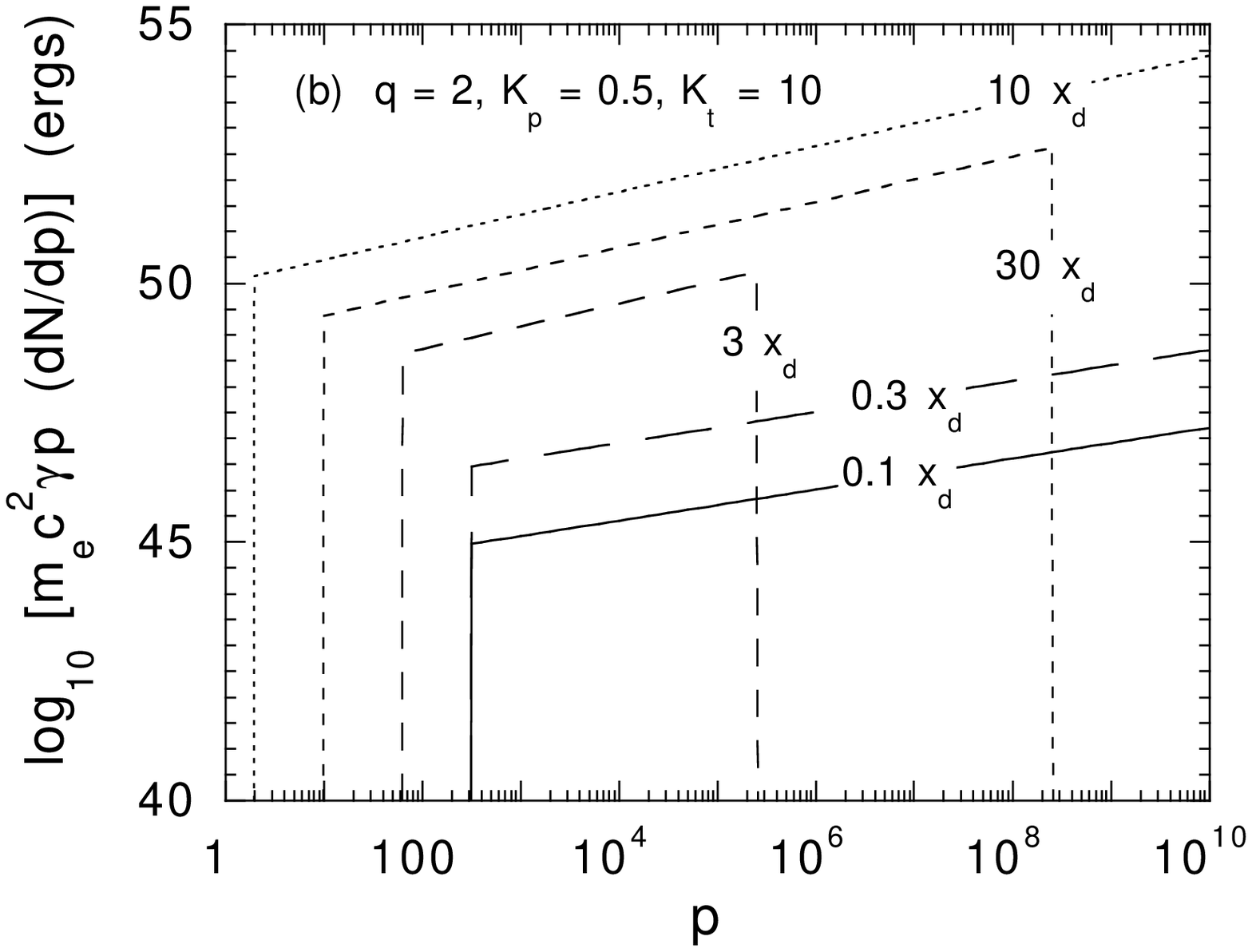}
\caption{Evolution of comoving particle distribution at different locations $x$ of the blast wave in units of the deceleration radius $x_d$ for the analytic result, equations (\ref{npx_ana1}) and (\ref{npx_ana1}), for a turbulence spectrum with $q=2$.  Parameters are $E_{54} = 1$, $n_0 = 100$ cm$^{-3}$, $\Gamma_0 = 300$, and $f_\Delta = 1/12$. In panel (a), $K_p = 0.5$ and $K_t = 2$. In panel (b), $K_p = 0.5$ and $K_t = 10$.
}
\end{figure}
\begin{figure}
\figurenum{4}
\epsscale{1.0}
\plottwo{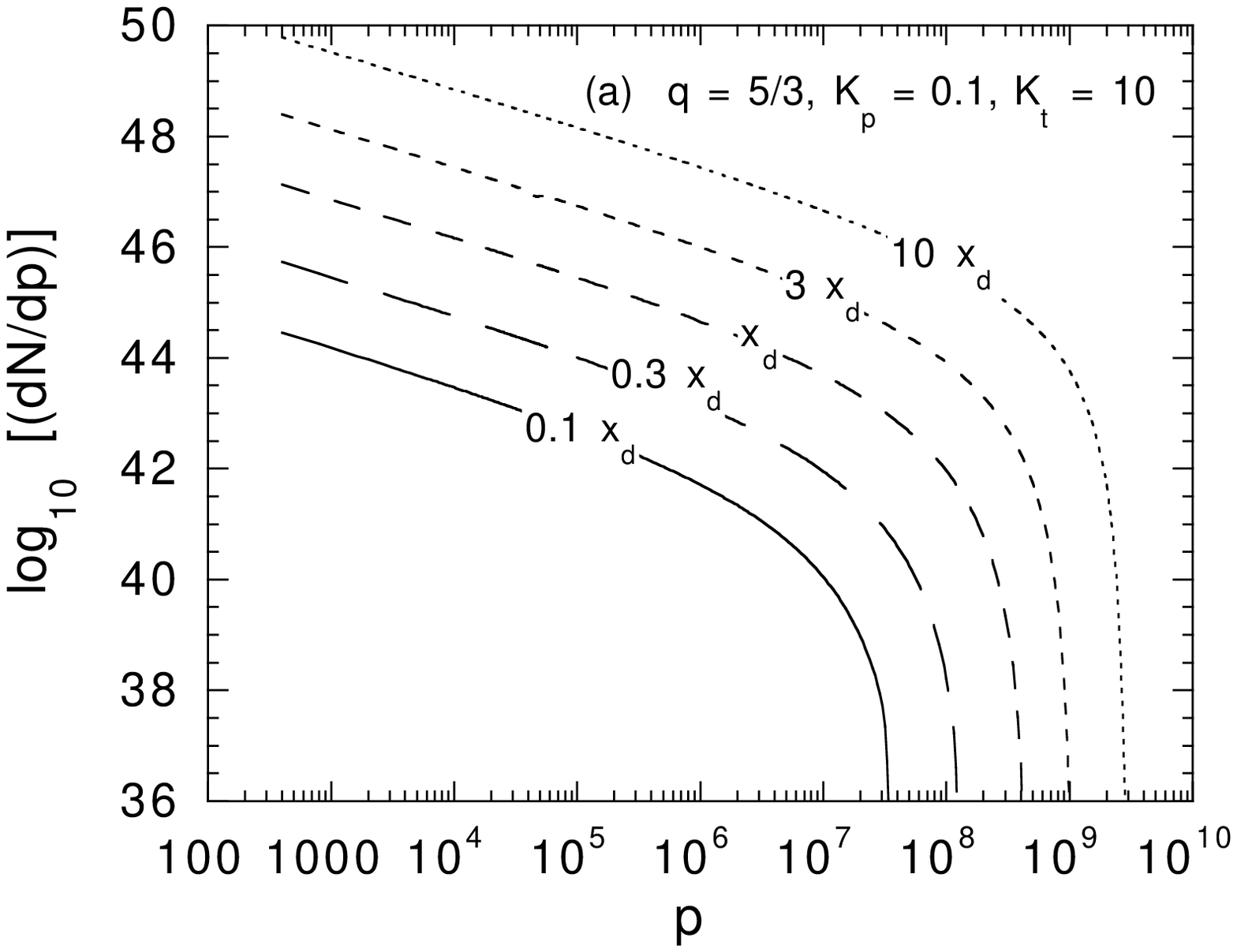}{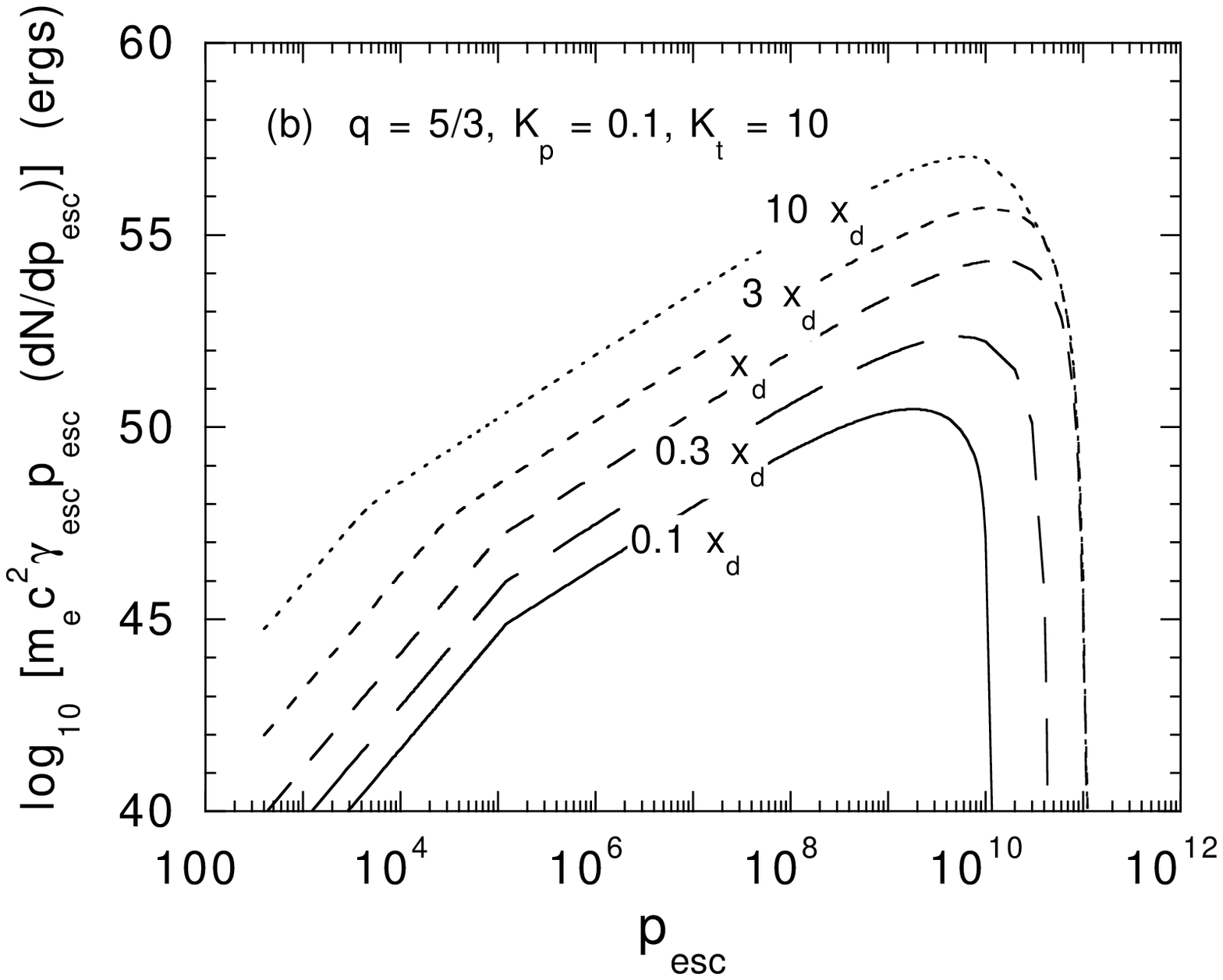}
\caption{Numerical calculation of the comoving particle distribution in panel (a), and the escaping particle distribution in panel (b) at different blast-wave locations.  Parameters are the same as in Fig.\ 3, except that $q = 5/3$, $K_p = 0.1$ and $K_t = 10$. 
}
\end{figure}
\begin{figure}
\figurenum{5}
\epsscale{1.0}
\plottwo{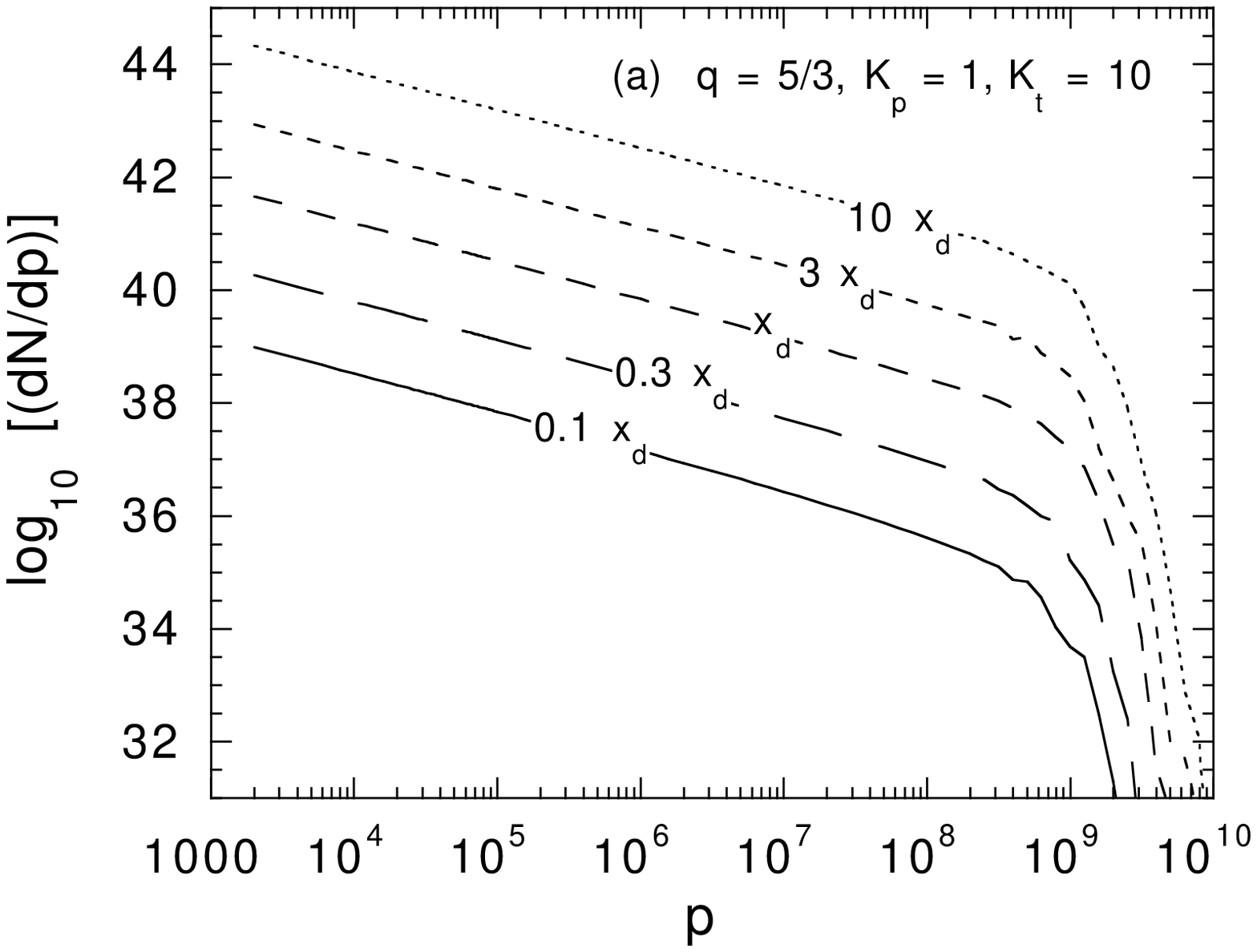}{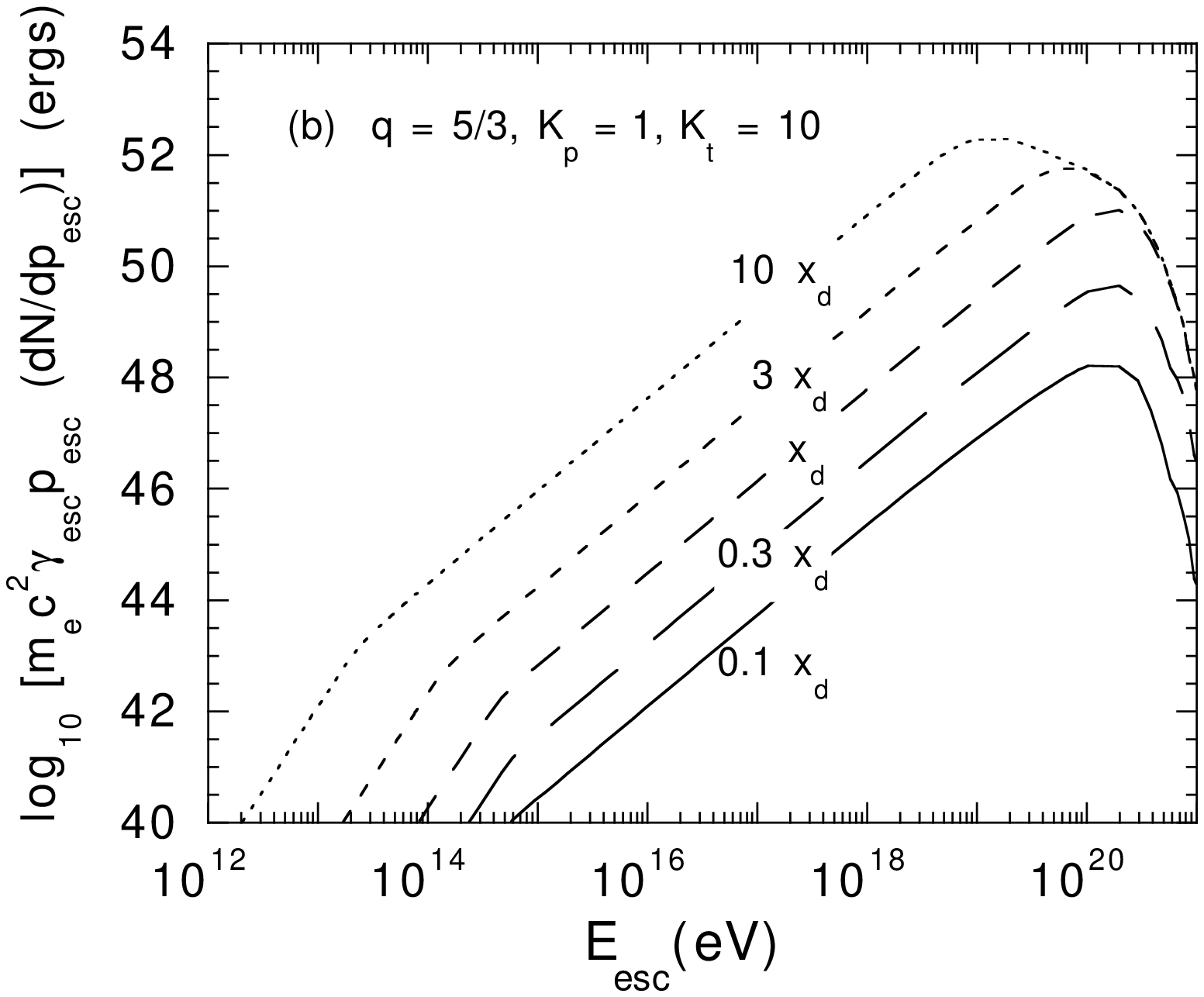}
\caption{Same as Fig.\ 4, except that  $K_p = 1$ and $K_t = 10$.
}
\end{figure}

\end{document}